\documentclass[%
 reprint,
 superscriptaddress,
 nofootinbib,
 amsmath,amssymb,
 aps,
 floatfix,
 showkeys,
]{revtex4-2}

\usepackage{graphicx}
\usepackage{dcolumn}
\usepackage{bm}
\usepackage{lineno,hyperref}
\modulolinenumbers[1]
\usepackage{multirow}
\usepackage{upgreek}
\usepackage[caption=false]{subfig}
\usepackage{siunitx}
\graphicspath{{Figures/}}

\begin{document}

\preprint{PRAB}

\title{Improved simulation of beam backgrounds and collimation at SuperKEKB}

\author{A.~Natochii}\email{natochii@hawaii.edu}\affiliation{University of Hawaii, Honolulu, Hawaii 96822, USA}
\author{S.~E.~Vahsen}\affiliation{University of Hawaii, Honolulu, Hawaii 96822, USA}
\author{H.~Nakayama}
\affiliation{High Energy Accelerator Research Organization (KEK), Tsukuba 305-0801, Japan}
\affiliation{The Graduate University for Advanced Studies (SOKENDAI), Hayama 240-0193, Japan}
\author{T.~Ishibashi}
\author{S.~Terui}
\affiliation{High Energy Accelerator Research Organization (KEK), Tsukuba 305-0801, Japan}

\date{\today}

\begin{abstract}
Mitigation of beam backgrounds via collimators is critical for the success of the Belle~II experiment at the SuperKEKB electron-positron collider. We report on an improved simulation methodology, which includes a refined physical description of the collimators and beam pipe, our first implementation of collimator tip scattering, and in which the existing beam particle tracking software has been embedded into a new sequential tracking framework. These improvements resolve longstanding discrepancies between measured and predicted Belle~II background levels, and significantly reduce the computing time required to optimize the collimation system in simulation. Finally, we report on collimator aperture scans, which confirm the accuracy of the simulation and suggest a new method for aligning the collimators.
\end{abstract}


\maketitle


\section{\label{sec:Introduction}Introduction}

The KEKB accelerator complex provided a world-record instantaneous luminosity of \SI{2.11e34}{cm^{-2}.s^{-1}} to the Belle experiment, which operated from 1999 through 2010 at the High Energy Accelerator Research Organization (KEK) in Japan. The upgraded Belle~II experiment served by the SuperKEKB electron-positron collider~\cite{REF2} seeks to achieve an unprecedented instantaneous luminosity of \SI{8.0e35}{cm^{-2}.s^{-1}} and to collect \SI{50}{ab^{-1}} of data in 10~years of stable operation. Recently, SuperKEKB achieved a new world record luminosity of \SI{2.4e34}{cm^{-2}.s^{-1}}~\cite{REF8}. In order to increase the luminosity by a factor of 40 compared to KEKB, the SuperKEKB design involves new beam optics that utilize a \textit{nano-beam scheme}~\cite{REF28} and higher beam currents of \SI{2.6}{A} and \SI{3.6}{A} for the electron and positron beam, respectively. These changes will also significantly increase the backgrounds from the machine. In particular, large beam losses near the interaction region where Belle~II is located, can adversely affect operational stability, quality of data, and detector longevity. The main goal of the collimation system is to protect the Belle~II detector and delicate machine components such as  superconducting magnets, while maintaining practical beam lifetimes, beam impedance, and injection performance.

In this article, we describe beam backgrounds caused by circulating beam particles interacting with their surroundings, such as the beam pipes, residual gas molecules, charges in the same bunch, and crossing beams. These interactions all involve elastic or inelastic scattering, which causes beam particles to deviate from their nominal trajectories. Some fraction of these particles end up being lost fully from the beam and hit the beam pipe, which produces showers of secondary background particles. Simulating machine-induced backgrounds is challenging and requires a good understanding of all processes causing beam losses. During the early commissioning stages of Belle~II and SuperKEKB, simulated and measured background rates differed by factors ranging from $\rm 10^{-2}-10^{3}$~\cite{REF3,REF26}. For collisions of the beam with residual gas molecules, discrepancies between simulation and measurement were expected, as details such as the pressure distribution and measured gas composition in the beam pipe had not been simulated. Subsequent work has steadily improved the understanding of the beam-gas component and will be reported in detail separately. For Touschek (intrabeam) scattering, however, the observed discrepancy was not expected, and hard to explain. At that time, simulated collimators would stop any incident particle hitting a collimator. One hypothesis to explain the Touschek discrepancy was, therefore, collimator leakage, where surviving particles scattering off of the collimator jaw (a process known as tip scattering) reach the interaction region. In this work, we finally resolve this Touschek discrepancy and show that its origin is indeed collimator leakage, however not via tip scattering as originally expected. We demonstrate instead that an improved simulation of the shape of each collimator leads to considerable changes in predicted background rates.

This article is structured as follows. In Section~\ref{sec:BeamBg}, we begin with an overview of the main SuperKEKB background processes, their measurement, and their simulation. We focus on measurements with a background monitoring system based on diamond detectors. For that reason, those detectors are described in detail in Section~\ref{subsec:RadiationMonitors}. 
Next, in Section~\ref{sec:CollimationSystem}, we review the collimation system. Given that the improved simulation of this system  had a particularly large impact and resolved the Touschek data/MC discrepancy for the electron beam, Section~\ref{subsec:ParticleTrackingAndImprovements} documents the detailed changes made to the simulation procedure, including the exact model for each individual collimator, which turned out to be a critical ingredient. Finally, the simulation is validated with collimator scans described in Section~\ref{sec:ReportOnValidationAndAchievements}, using diamond detectors to measure dose rates. The major conclusions and a summary of the research are provided in Section~\ref{sec:SummaryAndConclusions}. Details on improved beam-gas modelling, and more extensive validation measurements by all Belle~II sub-detectors, will be published separately in a forthcoming article.

\section{\label{sec:BeamBg}Beam background}

Here we give a brief overview of the major sources of beam-induced backgrounds at the SuperKEKB collider, their measurements using a radiation monitoring system, and the background simulation procedure.

\subsection{\label{subsec:BgSources}Background sources}

\begin{figure}[htbp]
\centering
\includegraphics[width=\linewidth]{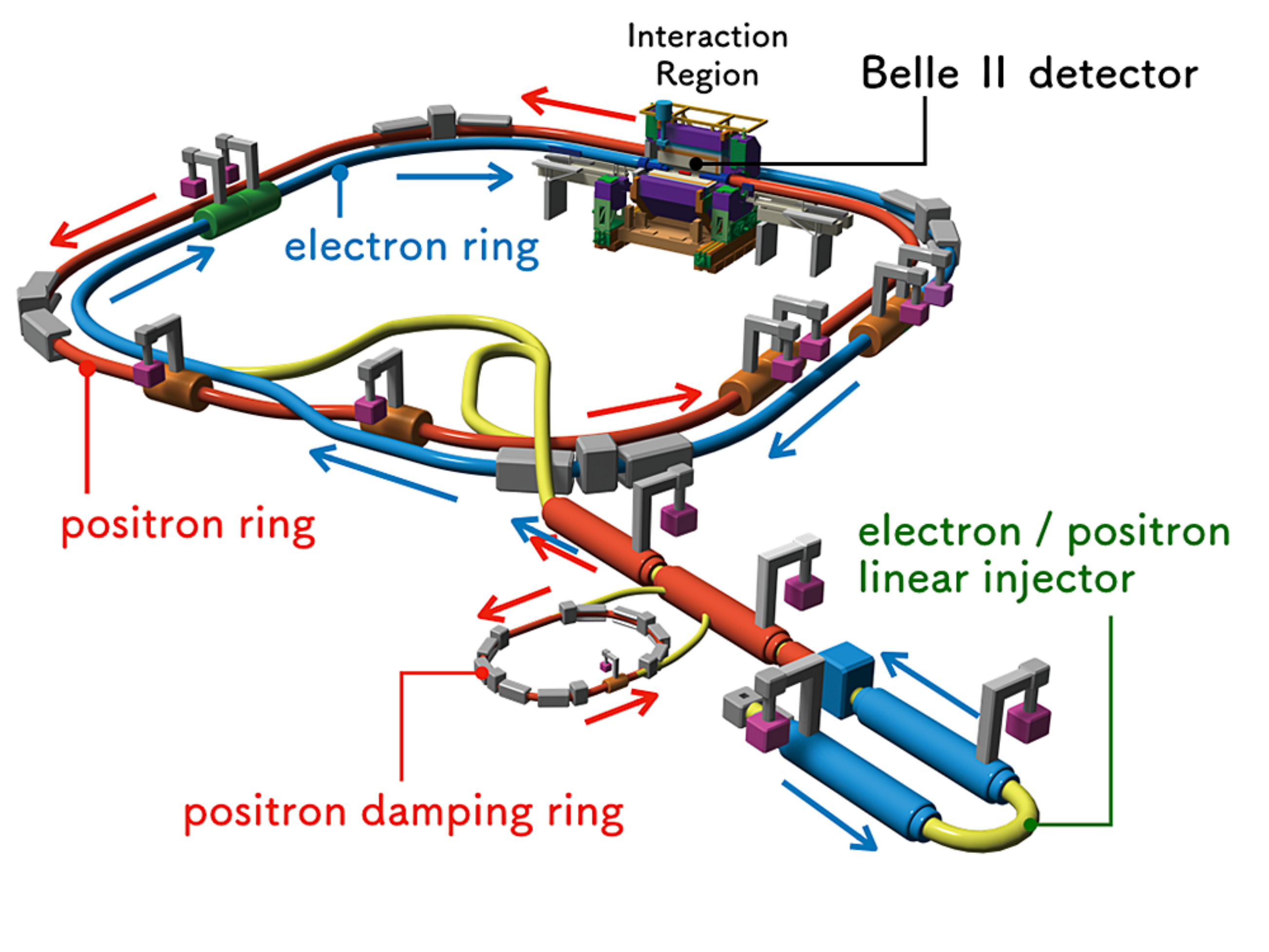}
\caption{\label{fig:fig1}Schematic drawing of the SuperKEKB collider complex, with Belle~II shown in the interaction region.}

\end{figure}

SuperKEKB is a high energy circular collider that consists of two rings, a \SI{7}{GeV} high energy electron ring (HER) and \SI{4}{GeV} low energy positron ring (LER). A comprehensive overview of the machine design is given in Ref.~\cite{REF2}. Figure~\ref{fig:fig1} depicts the SuperKEKB accelerator and related facilities, including the interaction region (IR) where the Belle~II detector is located. Belle~II extends approximately \SI{\pm 4}{m} from the interaction point (IP), where the two beams collide.

Compared to its predecessor, SuperKEKB is designed to operate with double the beam currents and twenty times smaller vertical beam sizes at the IP, both of which imply a significant increase of beam-induced backgrounds in the interaction region. The dominant expected background sources are 1) collision processes such as Radiative Bhabba scattering and two-photon processes, collectively referred to as luminosity backgrounds,  2) single-beam processes such as beam-gas scattering, including Bremsstrahlung and Coulomb scattering of beam particles with residual gas molecules, and Touschek scattering, which denotes Coulomb scattering between particles in the same bunch, 3) synchrotron radiation and 4) injection backgrounds induced by injected charges with large amplitudes of oscillation due to injection kicker errors~\cite{REF3}. 

We focus here primarily on Touschek and beam-gas scattering, as these two processes lead to off-orbit beam particles that form a beam halo, which then interacts with the machine aperture and leads to beam losses.

A set of countermeasures are used to mitigate beam-induced backgrounds. To suppress off-orbit particles, sets of 20 and 10 collimators with movable jaws are installed around the HER and LER, respectively. Vacuum scrubbing reduces the residual gas pressure in the beam pipe, thus suppressing beam-gas scattering. Heavy metal shields outside the IR beam pipe protect Belle~II against electromagnetic showers. A thin layer of gold on the inner surface of the beam pipe suppresses synchrotron radiation. More information about beam background mitigation at SuperKEKB can be found in Ref.~\cite{REF27}.

\subsection{\label{subsec:RadiationMonitors}Radiation monitors}

\begin{figure}[htbp]
\centering
\includegraphics[width=\linewidth]{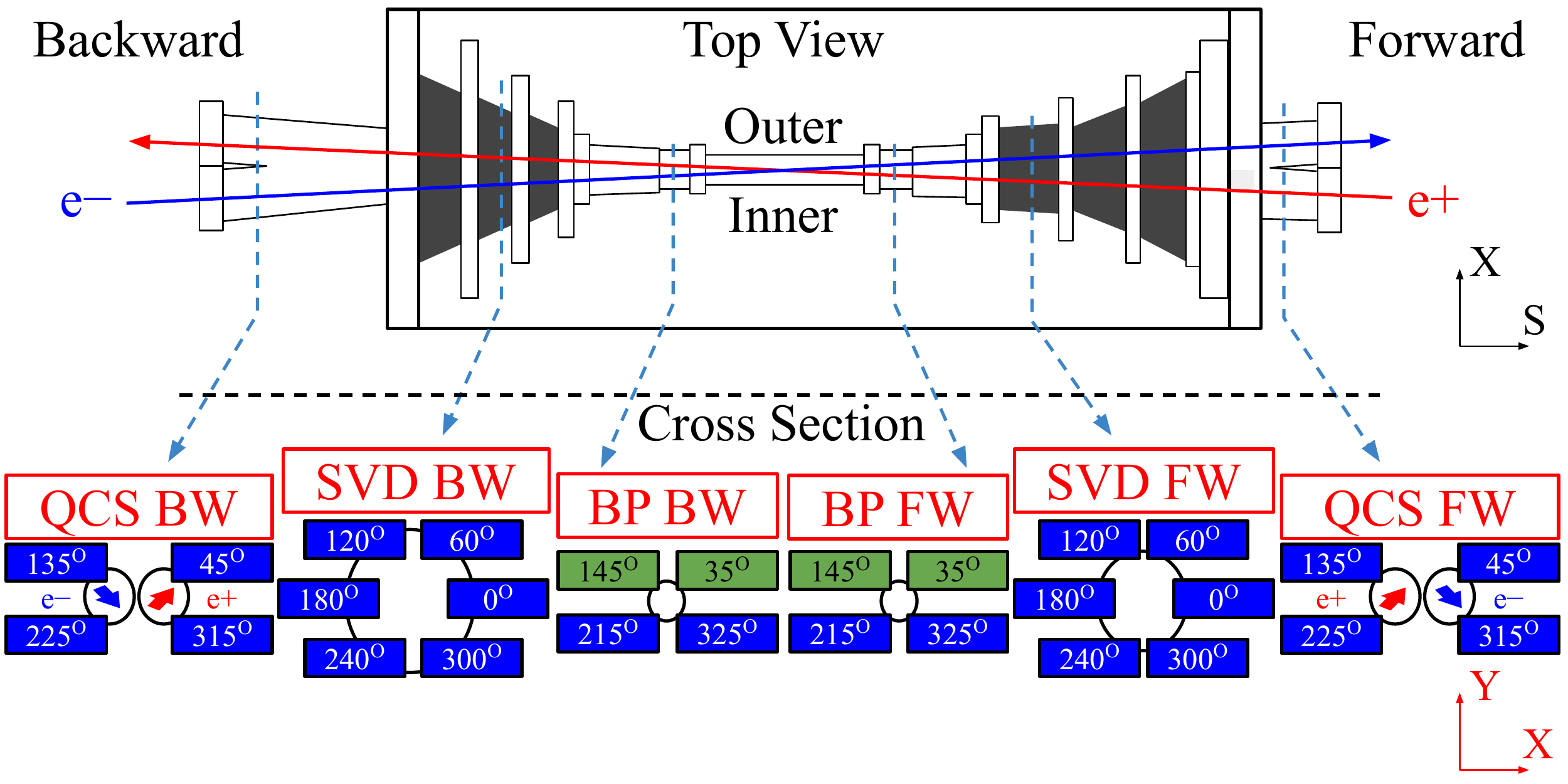}
\caption{\label{fig:fig9}Location of diamond detectors in the interaction region. Numbers in rectangles indicate each detector's azimuth angle. Blue and green rectangles indicate diamond detectors dedicated to dose rate monitoring at \SI{10}{Hz} and reserved for the beam abort function, respectively.}

\end{figure}

To ensure safe operation of the Belle~II detector, a dedicated background monitoring and beam abort system was installed at SuperKEKB. Its goal is to monitor the radiation dose rates around the IR beam pipe. The system consists of 28 diamond detectors mounted around the outside of the beam pipe in the interaction region, as shown in Figure~\ref{fig:fig9}. These detectors are grouped and named based on their location, as follows: four backward (\SI{-56.8}{cm}) QCS detectors (QCS BW); six backward (\SI{-27.4}{cm}) SVD detectors (SVD BW); four backward (\SI{-9.8}{cm}) and four forward (\SI[retain-explicit-plus]{+13.6}{cm}) Beam Pipe (BP) detectors; six forward (\SI[retain-explicit-plus]{+29.2}{cm}) SVD detectors (SVD FW); four forward (\SI[retain-explicit-plus]{+56.8}{cm}) QCS diamond detectors (QCS FW), where the numbers in parentheses are distances from the IP along the beam orbit.

The diamond sensor packaged into a detector unit is an artificial single-crystal produced by chemical vapor deposition (sCVD)~\cite{REF21}. Each sensor has a volume of \SI{4.5x4.5x0.5}{\milli\metre\cubed} and a mass of \SI{35.6}{mg}. 
For all 28 detectors dose rate data are read out at \SI{10}{Hz} for monitoring; buffer memories with data sampled at \SI{400}{kHz} are read out after beam aborts for post-abort analysis. The four upper BP diamond detectors (highlighted in green in Fig.~\ref{fig:fig9}) are enabled to generate beam abort request signals, when sums of \SI{400}{kHz} data in predefined moving time windows exceed programmed thresholds; they use a measurement range compatible with signals from large beam losses, with reduced sensitivity. The remaining 24 diamond detectors use a measurement range providing the highest sensitivity of about \SI{10}{\upmu rad/s}\footnote{\SI{1}{rad} = \SI{e-2}{Gy} is used as the dose unit in the text.} in dose rates read out at \SI{10}{Hz}, and are dedicated to monitoring. More information can be found in Ref.~\cite{REF24}. In the present article, monitoring diamond detectors are used to validate the improved beam background simulation as reported in Section~\ref{sec:ReportOnValidationAndAchievements}.


\section{\label{sec:CollimationSystem}Collimation system}

\begin{figure}[htbp]
\centering
\includegraphics[width=\linewidth]{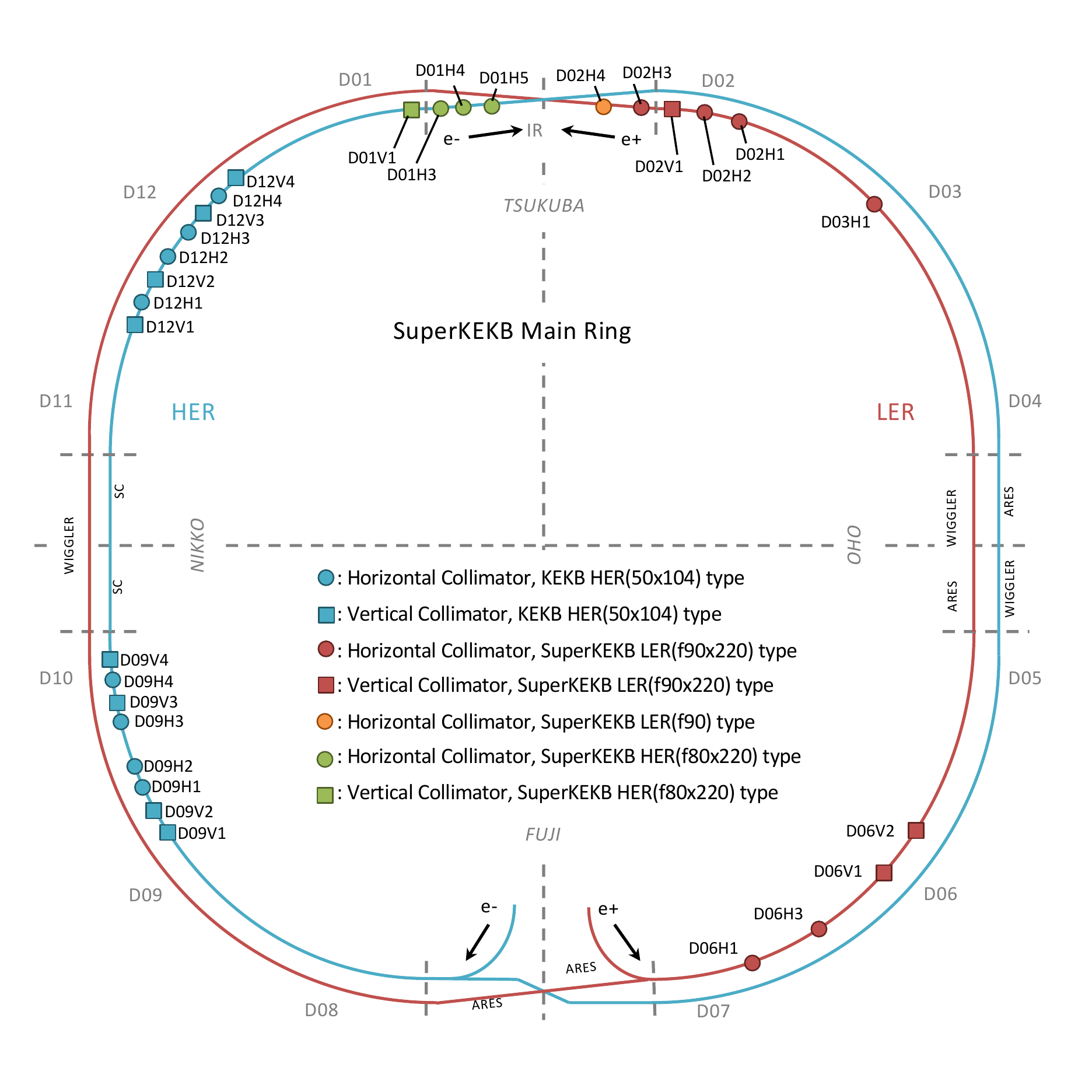}
\caption{\label{fig:fig2}Location of the SuperKEKB collimators as on June 2020. Vertical and horizontal collimator names contain the letters V and H, respectively. The storage ring is divided into twelve sections, which are referred to as D01 through D12.}
\end{figure}

At a high luminosity machine such as SuperKEKB, beam collimators are critical components necessary for reducing backgrounds in the interaction region to acceptable levels. They help to avoid superconducting magnet quenches and protect sensitive Belle~II electronics and sensors from stray beam particles and resulting background showers. Figure~\ref{fig:fig2} illustrates the current positions of SuperKEKB collimators. Each ring is divided into twelve sections, labeled D01 through D12. Sections D09 and D12 contain collimators inherited from KEKB, which have a jaw only on one side of the beam (\textit{KEKB-type}, Fig.~\ref{fig:fig31a}, \ref{fig:fig31b}) while all other collimators are new (\textit{SuperKEKB-type}, Fig.~\ref{fig:fig32a}, \ref{fig:fig32b}) and have two jaws.

\subsection{\label{subsec:Overview}Description of collimators}

\begin{figure}[htbp]
\centering
\includegraphics[width=\linewidth]{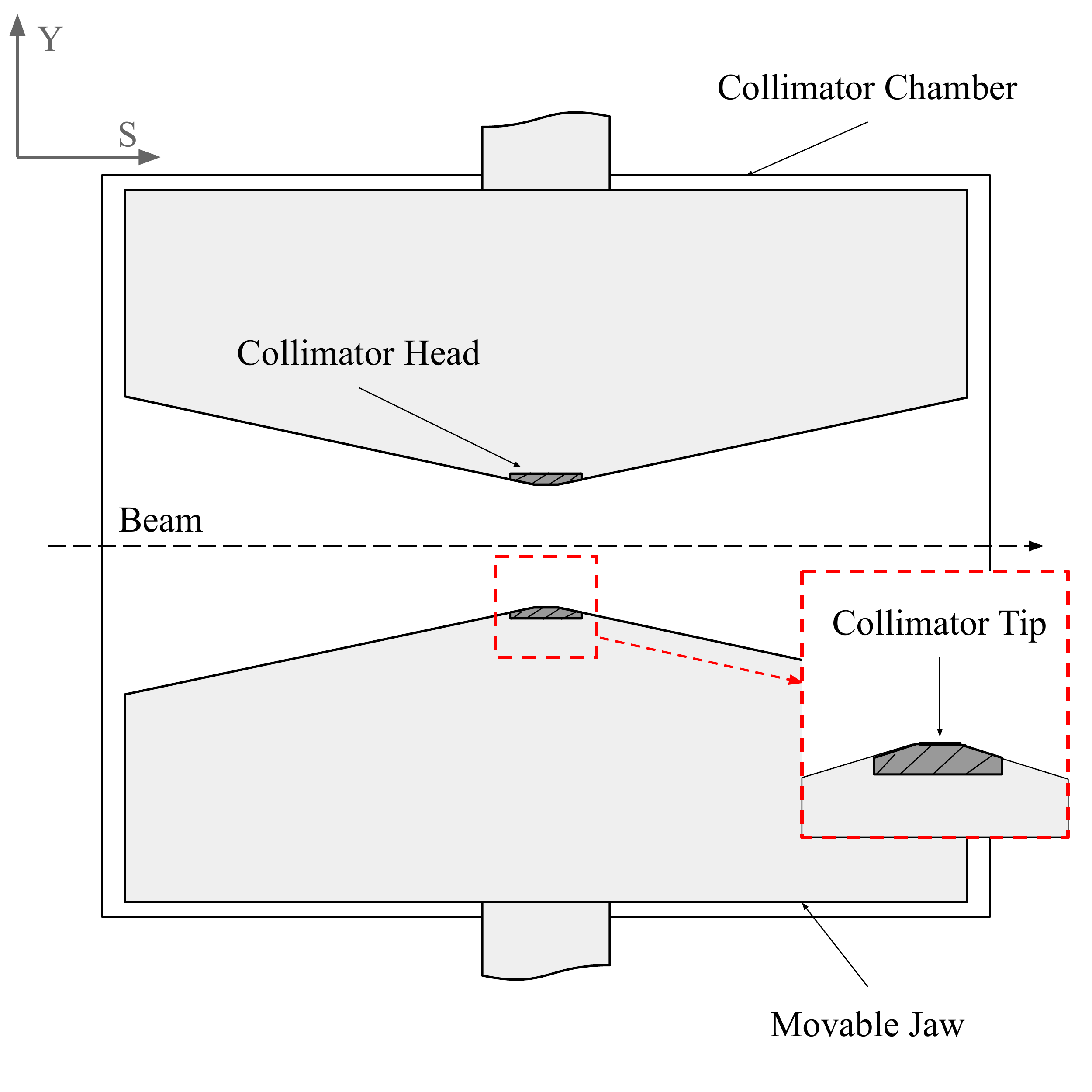}
\caption{\label{fig:fig23}A typical collimator structure: chamber, movable jaw, head, tip.}

\end{figure}

\begin{figure*}[htbp]
  \subfloat{%
    \begin{tabular}{c}
    \includegraphics[width=0.48\linewidth]{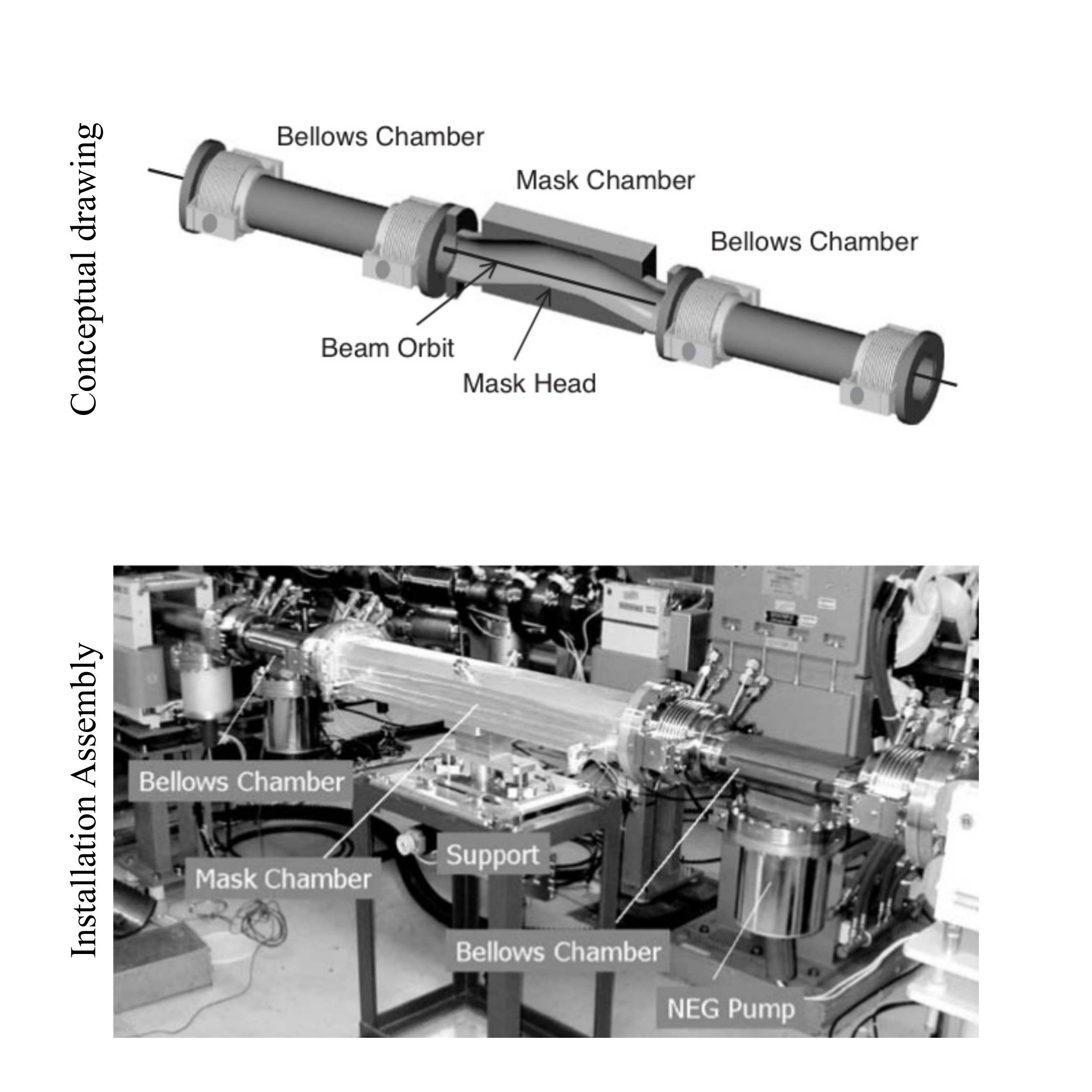}\\
    (a)\label{fig:fig31a}\\
    \end{tabular}
  }
\hfill
  \subfloat{%
    \begin{tabular}{c}
    \includegraphics[width=0.48\linewidth]{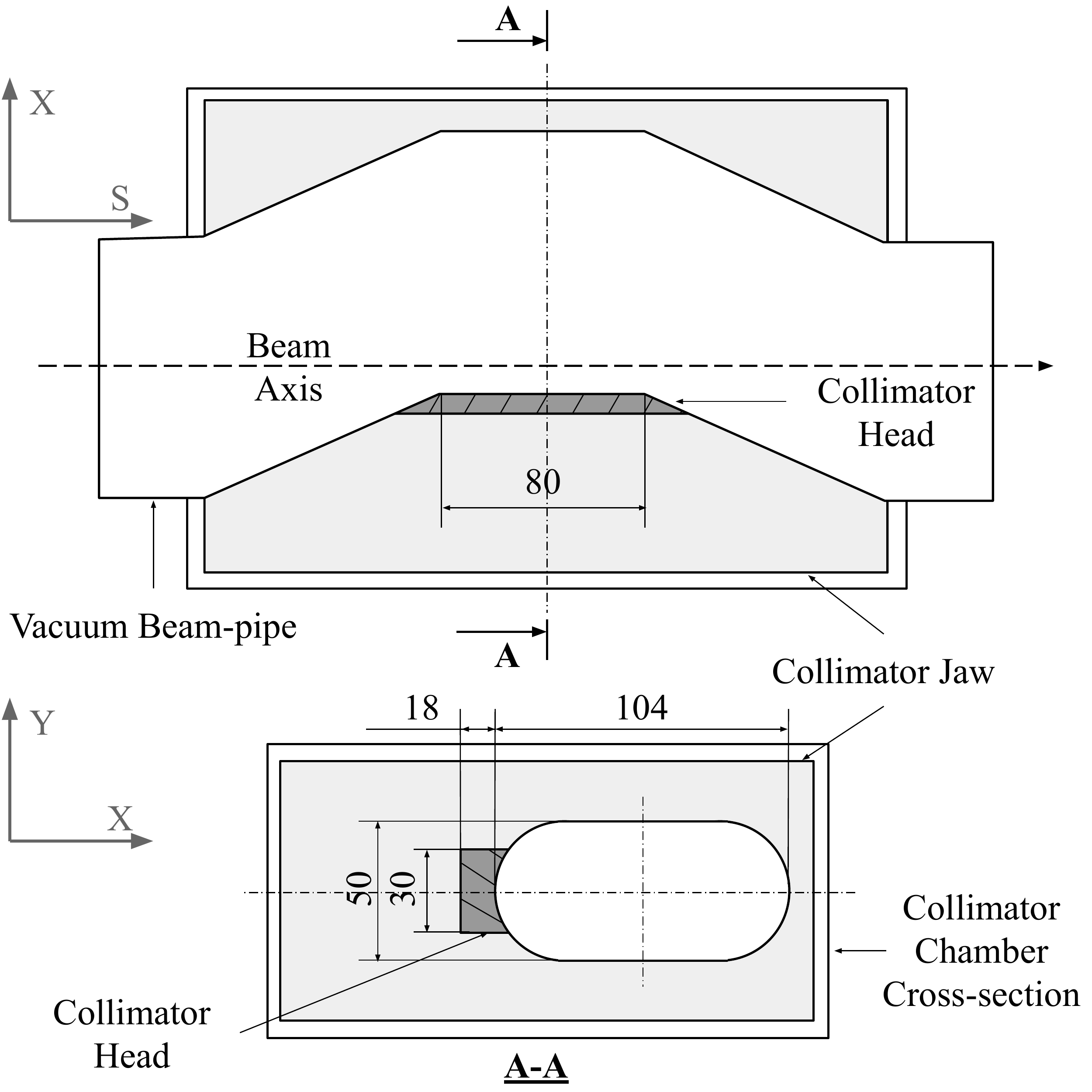}\\
    (b)\label{fig:fig31b}
    \end{tabular}
  }
  \caption{\label{fig:fig31}\textit{KEKB-type} collimators have a single jaw, on only one side of the beam. All dimensions are in mm. (a) Conceptual drawing and photo of the collimator assembly~\cite{REF12}. (b) Schematic drawing of the horizontal collimator chamber.}
    
\end{figure*}

\begin{figure*}[htbp]
  \subfloat{%
    \begin{tabular}{c}
    \includegraphics[width=0.48\linewidth]{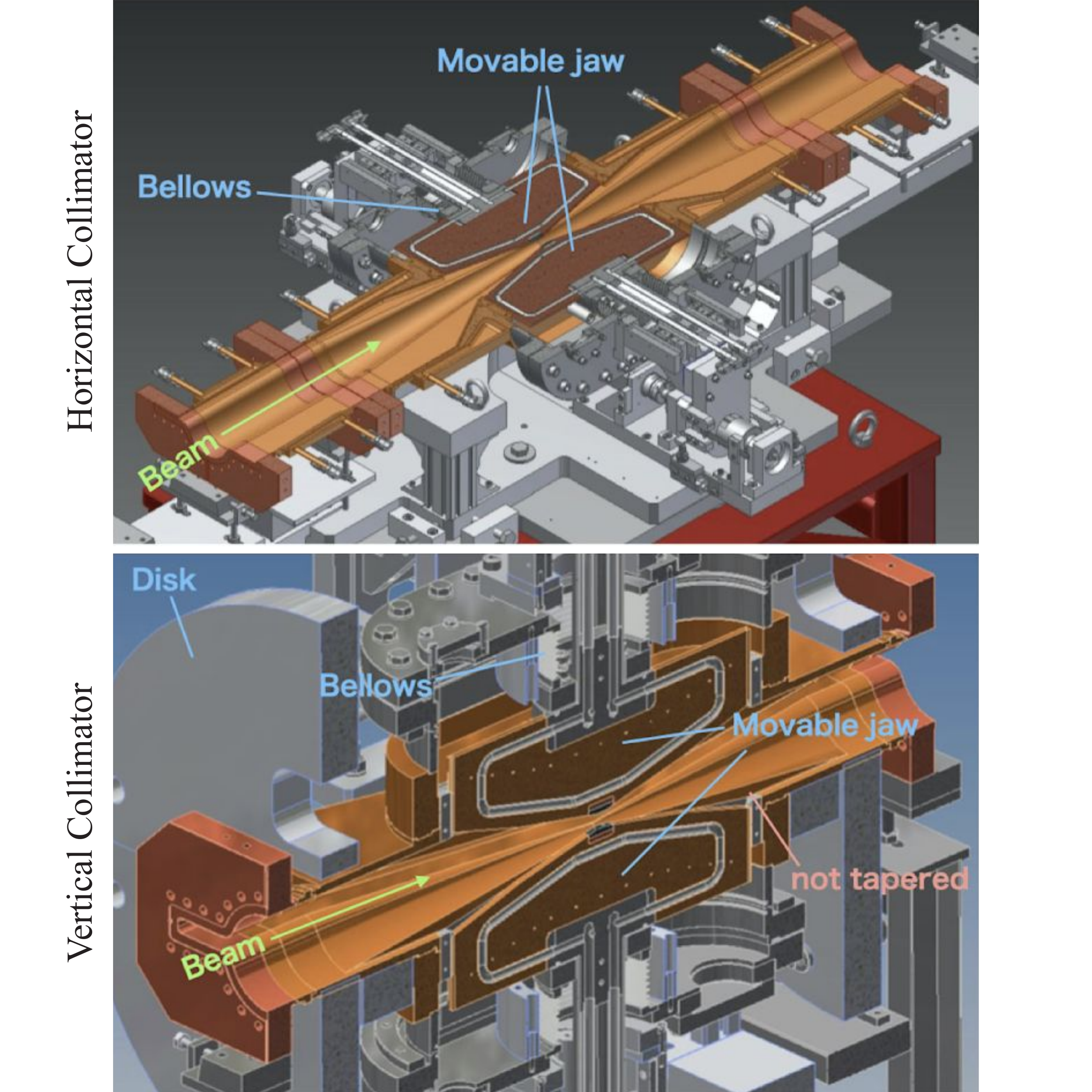}\\
    (a)\label{fig:fig32a}\\
    \end{tabular}
  }
\hfill
  \subfloat{%
    \begin{tabular}{c}
    \includegraphics[width=0.48\linewidth]{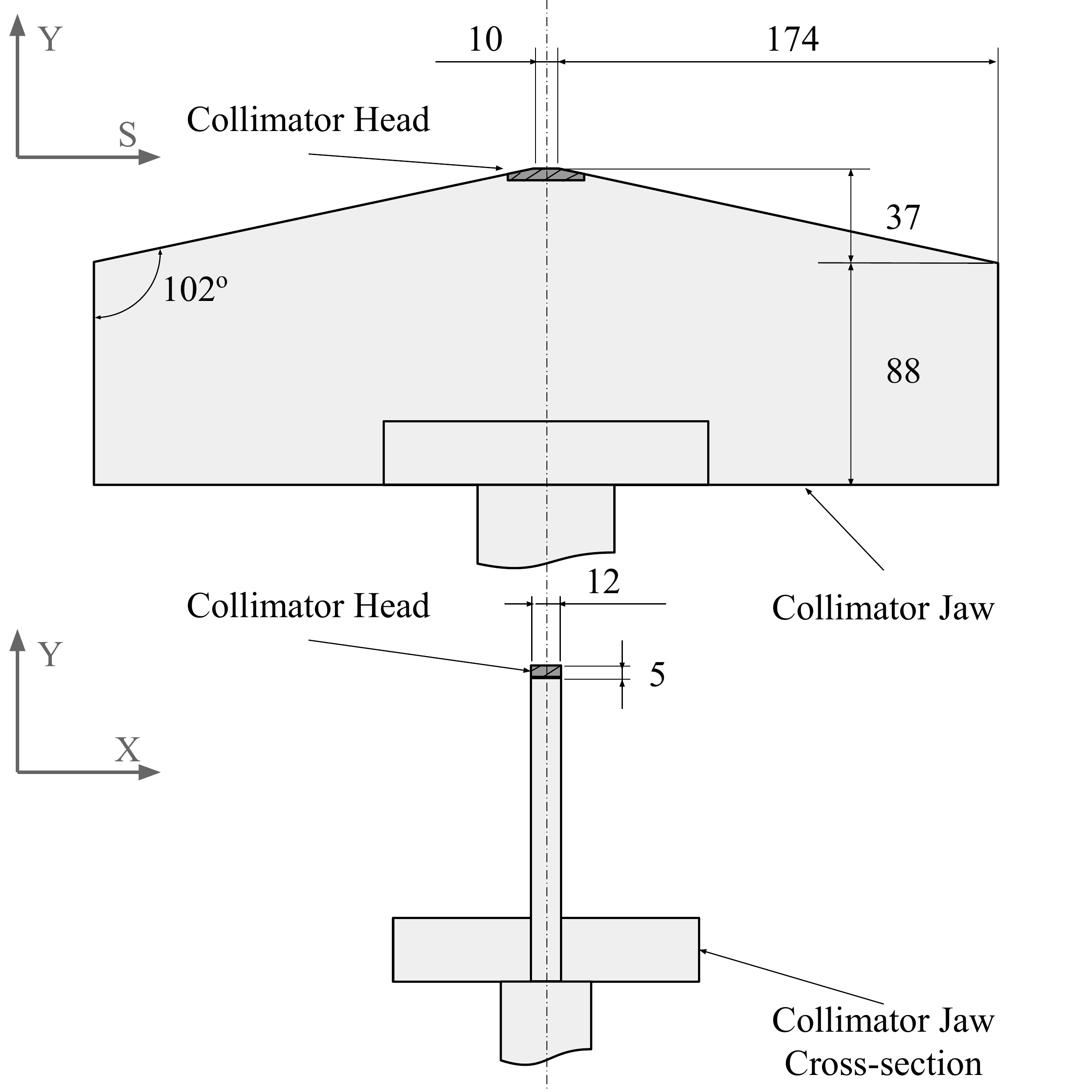}\\
    (b)\label{fig:fig32b}\\
  \end{tabular}
  }
  \caption{\label{fig:fig32}\textit{SuperKEKB-type} collimators have two jaws, one on each side of the beam. All dimensions are in mm. (a) Conceptual drawings of the collimator assembly~\cite{REF1}. (b) Schematic drawing of the collimator jaw.}
    
\end{figure*}

Before we proceed, let us first introduce the terminology used to describe the collimators, see Fig.~\ref{fig:fig23}: the collimator chamber is an element of the accelerator vacuum system which hosts movable jaws; the movable jaw is a tapered copper or aluminium block that absorbs stray beam particles far from the beam core; the collimator head is usually a high atomic number (high-\textit{Z}) material block used to collimate the particle beam, it is the part of the movable jaw closest to the beam core; the tip of the collimator jaw is the shortest surface of the head on the beam side, the collimator aperture is defined as the distance from the beam core to this surface of the head.

\begin{table*}[htbp]
\caption{\label{tab:tab1}Properties of the SuperKEKB collimators. The first column indicates the collimator type. Names of collimators with identical orientations and materials are grouped in the second column. The third column describes the orientation of each such group of collimators, while the fourth and fifth columns list the collimator jaw body and head materials. The radiation lengths (RLs) of these materials are~\cite{REF4}: RL(Cu) = \SI{14.4}{mm}, RL(Al) = \SI{89.0}{mm}, RL(W) = \SI{3.5}{mm}, RL(Ta) = \SI{4.1}{mm}, RL(Ti) = \SI{35.6}{mm}. The sixth column shows the length of the collimator tip in both millimetres and number of RLs.}
\begin{ruledtabular}
\begin{tabular}{ cccccc }
\multicolumn{2}{c}{Collimator} & \multirow{2}{*}{Orientation} & Body & Head & Tip length \\ \cline{1-2}
Type & Section/name &  & material & material & [mm]/[RL]\\
\hline
\multirow{4}{*}{\textit{SuperKEKB}}
            & D01-D06 & Hor.  & Cu &  W & 10.0 / 2.9 \\
            & D01-D02 & Vert. & Cu &  W & 10.0 / 2.9 \\
            & D06V1   & Vert. & Cu & Ta &  5.0 / 1.2 \\
            & D06V2   & Vert. & Cu & Ta & 10.0 / 2.4 \\
\hline
\multirow{2}{*}{\textit{KEKB}}
			& D09-D12 & Hor.    & Cu & Ti & 80.0 / 2.2 \\
			& D09-D12 & Vert.   & Al & Ti & 40.0 / 1.1 \\
\end{tabular}
\end{ruledtabular}
\end{table*}

Here it is worth mentioning the main differences between collimator types used in the collider, which are summarized in Table~\ref{tab:tab1}. All \textit{SuperKEKB-type} collimators (vertical and horizontal), with the exception of D06V1 and D06V2, have a tungsten (W) head with a \SI{10}{mm} long tip. The vertical collimator D06V1, which was installed during the 2019-2020 winter shutdown, has a tantalum (Ta) head with a shorter, \SI{5}{mm}, tip, while the vertical collimator D06V2 has a tantalum head with a \SI{10}{mm} tip. The jaw of \textit{SuperKEKB-type} collimators (i.e., disregarding the head made of high-\textit{Z} material) is made of copper (Cu). While \textit{SuperKEKB-type} collimators have heads that are \SI{12}{mm} wide transverse to the beam direction, \textit{KEKB-type} collimators have race-track-shaped cross-sections and much wider heads. The jaws of \textit{KEKB-type} horizontal and vertical collimators are made of Cu and aluminum (Al), respectively. Their heads are made of titanium (Ti) with a \SI{80}{mm} tip length for horizontal and \SI{40}{mm} tip length for vertical collimators. Each movable jaw is equipped with a linear variable differential transformer (LVDT) with a position accuracy better than \SI{30}{\upmu m} and \SI{80}{\upmu m} while the repeatability is about \SI{50}{\upmu m} and \SI{100}{\upmu m} for \textit{SuperKEKB-type} and \textit{KEKB-type} collimators, respectively. The alignment precision of the collimator chamber 
with respect to the nearest quadrupole magnet, which defines the transverse beam position, varies from \SI{200}{\upmu m} up to \SI{500}{\upmu m} for vertical \textit{SuperKEKB-type} collimators. For horizontal \textit{SuperKEKB-type} and all \textit{KEKB-type} collimators the precision varies from \SI{500}{\upmu m} up to \SI{1000}{\upmu m}. This uncertainty may cause an unintended offset between the reference of the collimator head position and the beam core.

\subsection{\label{subsec:OptimizationProcedure}Collimation system optimization}

Careful adjustment of collimators is necessary to reduce experimental backgrounds to acceptable levels for stable operation of Belle~II. SuperKEKB thus follows a \textit{collimation system optimization} procedure where an operator manipulates collimator jaw positions to fulfil a set of requirements. The aim is to find suitable positions for each collimator jaw that both reduce backgrounds and maintain adequate beam lifetimes. Since SuperKEKB maintains constant beam currents by performing continuous top-up injections, it is also necessary to achieve a high injection efficiency, which is defined as the ratio of stored to injected charge. A lower injection efficiency requires a higher time-averaged (net) injection rate, which in turn leads to a reduction in Belle~II data taking efficiency, as Belle~II uses a Level-1 trigger veto synchronized with injections. Aggressively closing the collimators reduces the machine's physical aperture, which is determined by the beam pipe's cross-section, including collimators. This results in lower injection efficiency, as well as higher local losses at collimators, leading to undesirable activation, heating, and potential collimator damage. On the other hand, a wide open configuration leads to an increase of injection and storage backgrounds. Therefore, the collimation system optimization requires a compromise between (i) maintaining good injection performance and beam lifetime, both of which favor wider collimators, and (ii) reducing backgrounds in Belle II, which favors narrower collimators. Relying on a well-known algorithm based on phase advance analysis, which is also known as \textit{betatron collimation}~\cite{REF14}, operators aim to find the most optimal aperture of collimators. A half-integer phase-advance between the IR and a given collimator (see Fig.~3 in Ref.~\cite{REF1}) provides an effective reduction of IR losses and hence of Belle~II backgrounds. This optimization procedure is performed at low beam currents to avoid accidental collimator head damage, and requires an accurate tuning of the radial position of each jaw, one by one, by monitoring interaction region losses and injection efficiency.

\section{\label{subsec:ParticleTrackingAndImprovements}Particle tracking simulation and recent improvements}

The beam background simulation starts with a multi-turn particle tracking framework named Strategic Accelerator Design (SAD)~\cite{REF5}. SAD is initialized with a machine optics file, which describes the collider configuration and all collimator apertures during a dedicated measurement. The program tracks scattered particles through the machine lattice. 
Particles lost in the interaction region are exported from SAD to Geant4~\cite{REF29,REF30,REF31} for simulation of the detector response. Geant4 scripts realistically describe the geometry of the detector and produce digitized observables (e.g., dose rate) to be compared against the experimental data. 


Here we focus on two single-beam processes, beam-gas scattering and Touschek scattering, and their interaction with collimators. These two processes are the dominant sources of beam backgrounds at low luminosities. Luminosity backgrounds, which we expect to dominate at higher luminosities, are not affected by the collimators.

In the Belle~II collaboration, the ratio of measured to MC simulated rates for a given background component is colloquially referred to as the "data/MC" ratio. Such ratios are obtained from dedicated single-beam and luminosity background studies conducted at SuperKEKB. A description of such measurements and of the data/MC calculation procedure can be found in Ref.~\cite{REF3}. The ultimate goal of the simulation is to reach data/MC~$\approx$~1, which would demonstrate that our understanding of the machine is sufficient to predict the background behaviour at different beam currents and optics. In the present paper, we focus on the MC side, and explain in detail how the simulation has been improved.

\subsection{\label{subsec:ExistSimInfrastructure}Existing simulation infrastructure}

The production of an MC sample for a given machine configuration, including beam optics, collimation system settings, and beam currents, involves two main steps. A multi-turn beam simulation tracks scattered beam particles through the lattice of the accelerator. This involves the software package SAD, which is described below. Beam particles lost near the IR are tracked in Geant4, which predicts secondary showers due to interaction with material, and the subsequent response of Belle~II sub-detectors and diamond beam background monitors. The present article is concerned mainly with the first step, which we will now describe in more detail.

SAD is a simulation framework for accelerator design and particle tracking, which has been developed at KEK since 1986. Single-beam particle tracking starts by defining about 500 equidistant \textit{scattering} positions along the beam axis. The coordinate advancing along the nominal beam particle trajectory is named \textit{S}. At each scattering position, the scattering of electrons or positrons is simulated. A vertical offset with respect to the beam center is applied to each sextupole magnet position. These offsets generate machine errors referred to as X-Y coupling. The generated offsets are normally distributed with a standard deviation of \SI{50}{\upmu m} around the beam center. We assume particle bunches are 3D Gaussian distributions with a volume of $\sigma_{\rm X}\times\sigma_{\rm Y}\times\sigma_{\rm S}$, where $\sigma_{\rm X/Y} = \sqrt{\beta_{\rm X/Y}\times\varepsilon_{\rm X/Y}}$ is the standard deviation of the normal distribution in the XY-plane, while $\sigma_{\rm S}$ is the standard deviation along the beam axis and hence is the length of the bunch, $\beta$ is the beta function at the scattering position, and $\varepsilon$ is the emittance of the beam. Randomly distributed particles are generated at each scattering position with a modified momentum and a statistical weight factor, both calculated based on beam-gas and Touschek scattering theories~\cite{REF2}.

\begin{figure*}[htbp]
  \subfloat{%
  \begin{tabular}{c}
    \includegraphics[width=0.48\linewidth]{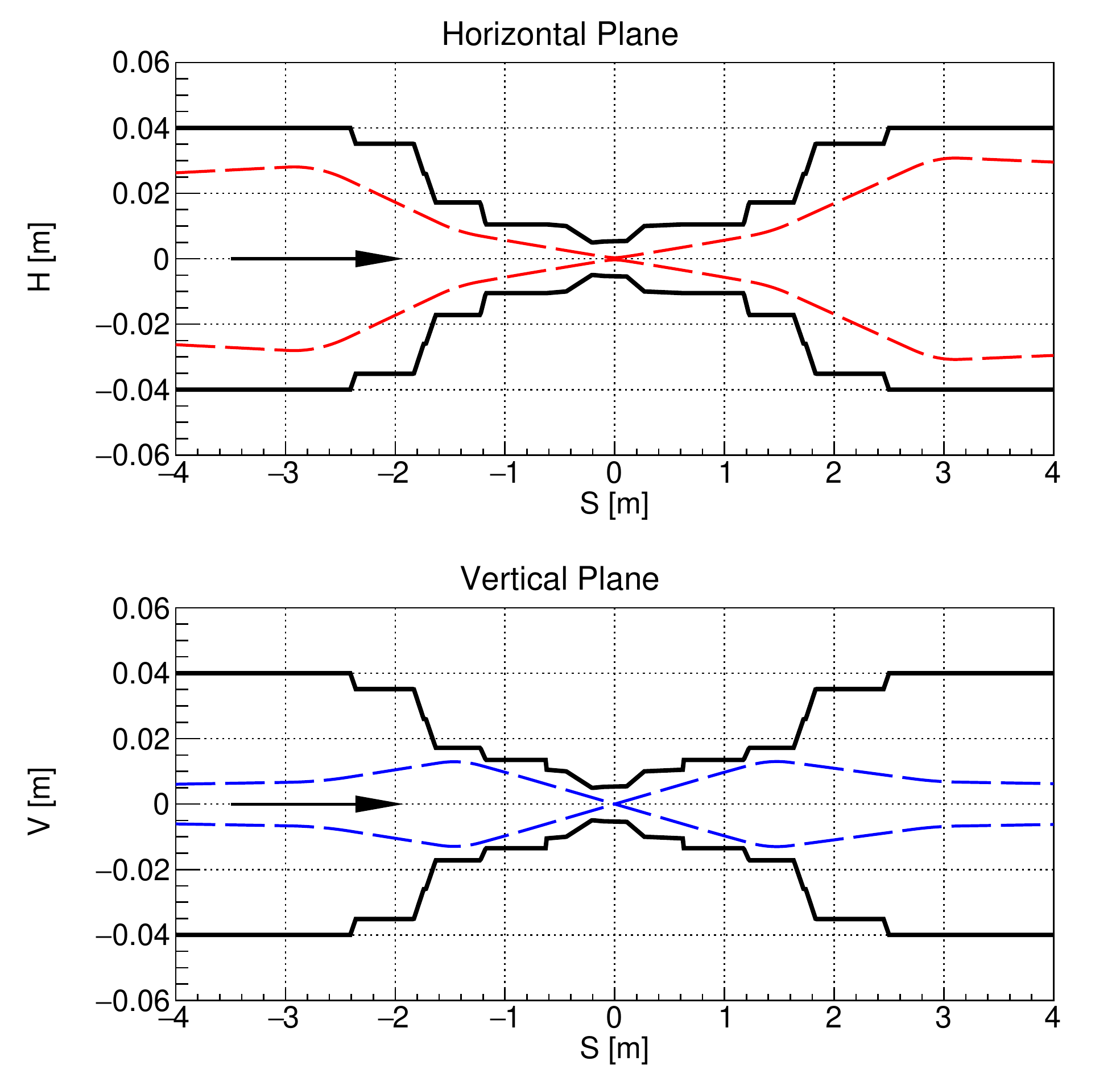}\\
    (a)~HER\label{fig:fig5a}\\
  \end{tabular}
  }
  \hfill
  \subfloat{%
  \begin{tabular}{c}
    \includegraphics[width=0.48\linewidth]{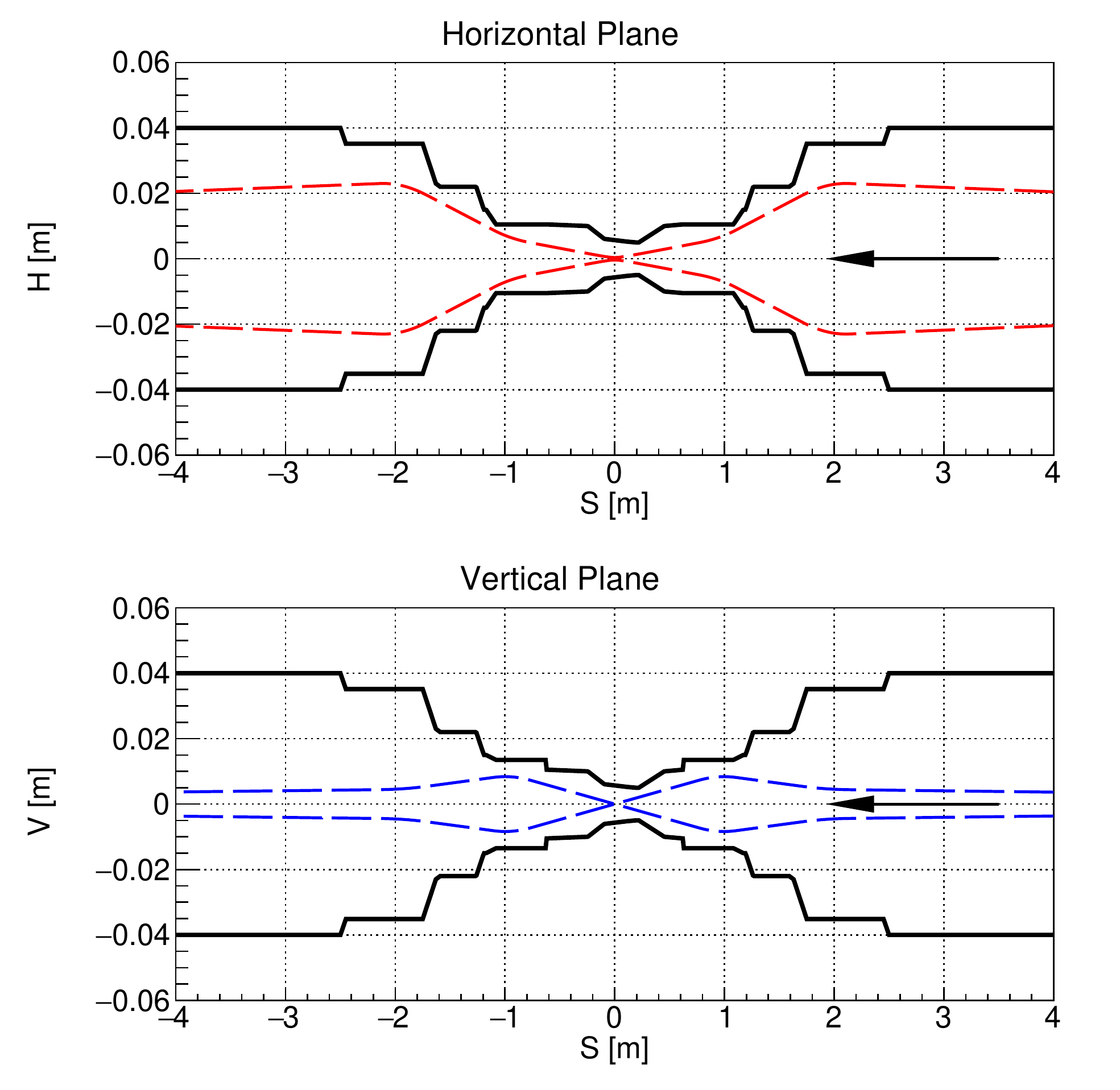}\\
    (b)~LER\label{fig:fig5c}\\
  \end{tabular}
  }
  \caption{\label{fig:fig5}Physical aperture of the HER (a) and LER (b) beam pipes in the interaction region, as implemented in SAD. Colored lines represent the beam envelope, which is limited by collimators. Arrows indicate the beam direction. The beam optics simulated are those of June 2020: a beta function at the IP of $\beta^{*}_{\rm H/V} = \SI{60/0.8}{mm}$ and crab sextupoles turned on for a so-called Crab Waist collision scheme~\cite{REF18}.}
    
\end{figure*}

We recently improved the IR beam pipe geometry in SAD, so that it agrees better with the experimental reality and the shape used in the Belle~II Geant4 model, see Fig.~\ref{fig:fig5}. When running SAD, a set of input parameters is specified. These include $Z_{\rm eff}$, the effective atomic number of the residual gas in the beam pipe. We assume $Z_{\rm eff}=7$, mostly due to CO molecules~\cite{REF11}. The number of simulated machine turns is usually set to 1000. We also enable synchrotron radiation and acceleration by radiofrequency cavities in the SAD simulation. Beam particles that cross the inner beam pipe surface or hit a collimator aperture are denoted as lost beam particles, and these are not tracked further by SAD. Instead, the simulation passes lost particles into Geant4, which simulates their interaction with surrounding materials and detectors.

\begin{figure*}[htbp]
  \subfloat{%
  \begin{tabular}{c}
    \includegraphics[width=0.48\linewidth]{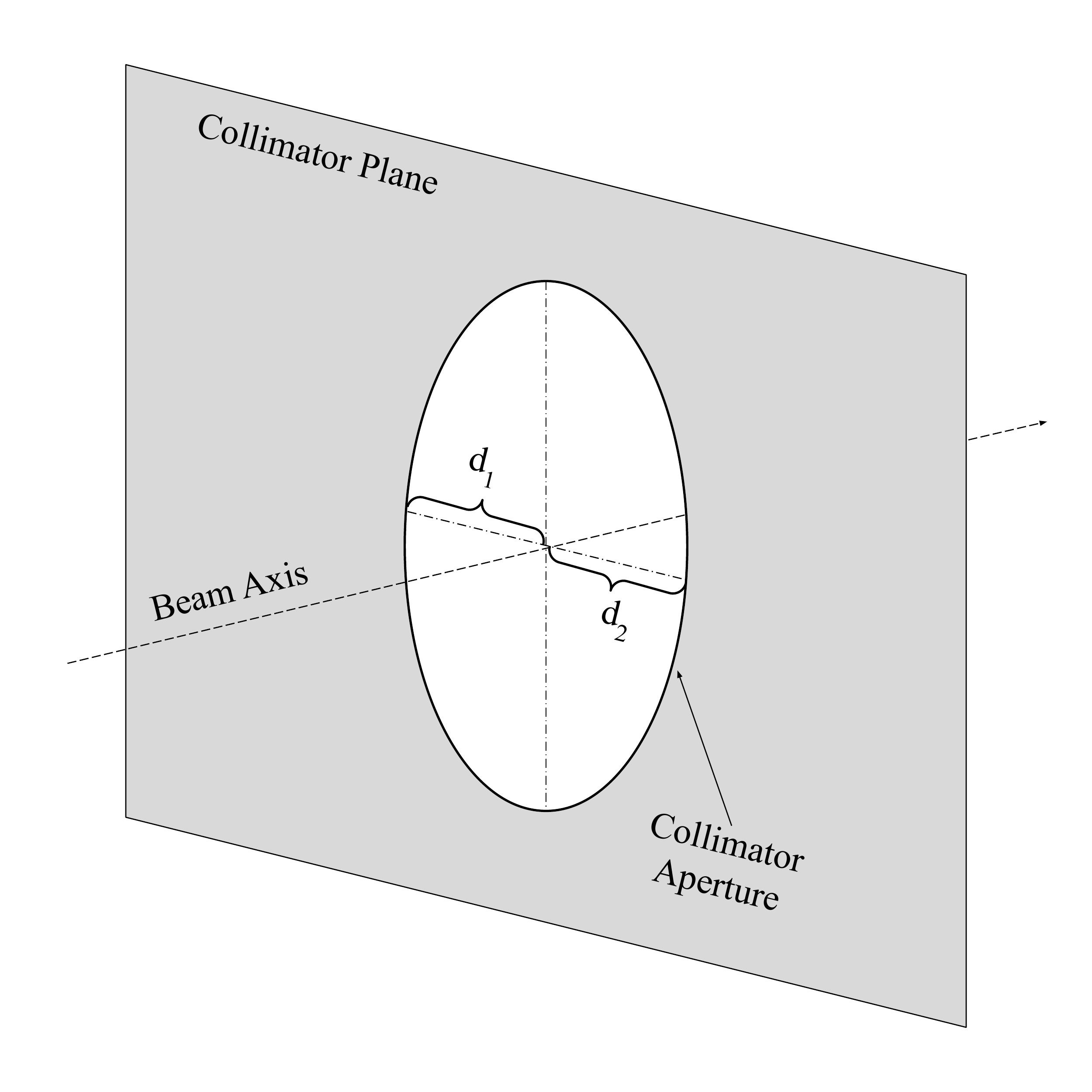}\\
    (a)~Horizontal\label{fig:fig4a}\\
  \end{tabular}
  }
  \hfill
  \subfloat{%
  \begin{tabular}{c}
    \includegraphics[width=0.48\linewidth]{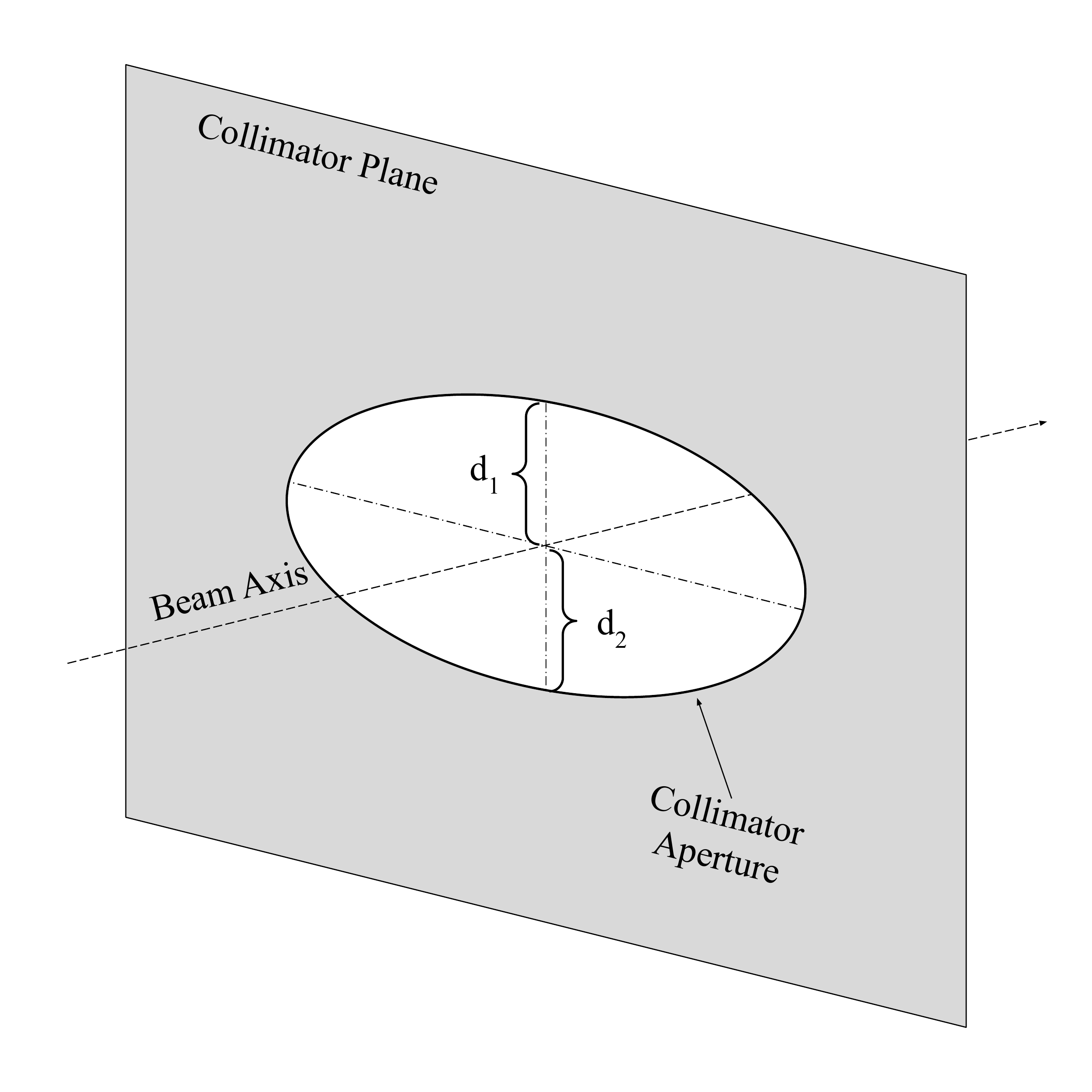}\\
    (b)~Vertical\label{fig:fig4c}\\
  \end{tabular}
  }
  \caption{\label{fig:fig4}Default horizontal (a) and vertical (b) collimator models in SAD. The ellipse half-axes $d_{\rm 1}$ and $d_{\rm 2}$ represent the distances from the beam axis to the left/top and right/bottom tip of the collimator, respectively. In each case, the second half-axis (vertical for horizontal collimator and horizontal for vertical collimator) is set equal to the beam pipe radius \SI{45}{mm}. Particles hitting the grey area are classified as lost.}
    
\end{figure*}

Historically at KEK, the particle tracking simulation was organised so that all lost particles were collected after a given number of machine turns in SAD. Only elliptical shapes, rectangular shapes, and a mix of both shapes could be used to describe collimators. By default, collimators were implemented as infinitely thin planes that absorb all particles outside the elliptical aperture defined by the half-axes $d_{\rm 1}$ and $d_{\rm 2}$, as shown in Fig.~\ref{fig:fig4}.

To predict the losses in the IR with reasonable statistics, we typically generate 100~particles per scattering position per ring, for every single-beam scattering process (i.e., Coulomb, Bremsstrahlung, and Touschek). The KEK Central Computer system~\cite{REF32, REF33} enables us to run more than 1000 simulations in parallel, thereby improving statistics and reducing computation time. Each final tracking simulation sample contains about $10^{6}-10^{8}$ generated particles per ring.

SAD stores beam particle coordinates at the end of lattice elements that have an aperture narrower than the transverse particle position. Such particles are considered lost, traced back to the inner beam pipe surface, and passed to Geant4, which is used to simulate interactions with the beam pipe walls.

\subsection{\label{subsec:SequentialTracking}New sequential tracking algorithm}

To augment the functionality of the particle tracking in SAD, we developed a new  \textit{sequential tracking algorithm}, which reports on stray beam particles turn-by-turn. The idea of this method is as follows: after beginning particle tracking, SAD stops at each collimator and stores the 6D coordinates of each beam particle. Afterwards, the simulation continues tracking until the next collimator. This process continues, with the algorithm sequentially stopping at collimators until 1000 machine turns have been reached, upon which the routine is terminated. At the end of the code execution, there is now not only information about lost particles, but also a history of the beam dynamics across all simulated machine revolutions. This straightforward algorithm gives many benefits and allows for the study of beam evolution. Below, we discuss some of the essential features resulting from the sequential tracking algorithm.

\subsection{\label{subsubsec:CollimatorProfile}Corrected collimator profiles}

 Collimators used at KEKB all had heads with race-track shaped cross sections, see Fig.~\ref{fig:fig31}. A fairly accurate elliptical approximation to this was therefore implemented as the default collimator shape in SAD, see Fig.~\ref{fig:fig4}. At SuperKEKB, however, only collimators D09 and D12 are of the original \textit{KEKB-type}, while all the others are of the new \textit{SuperKEKB-type}. These newer collimators have two opposite jaws with a rectangular head shape of \SI{12}{mm} width transverse to the beam axis direction (Fig.~\ref{fig:fig32}). To improve the simulation accuracy, we implemented a more realistic \textit{SuperKEKB-type} collimator shape in the SAD particle tracking scripts. This was enabled by the sequential tracking technique introduced above.

At the tracking stage, the simulation program uses information about the particle distribution at each collimator to define which part of the beam interacts with a given collimator. Particles outside the edge of the jaw are considered as not absorbed, and the routine continues to track them around the ring.

\begin{figure}[htbp]
\centering
\includegraphics[width=\linewidth]{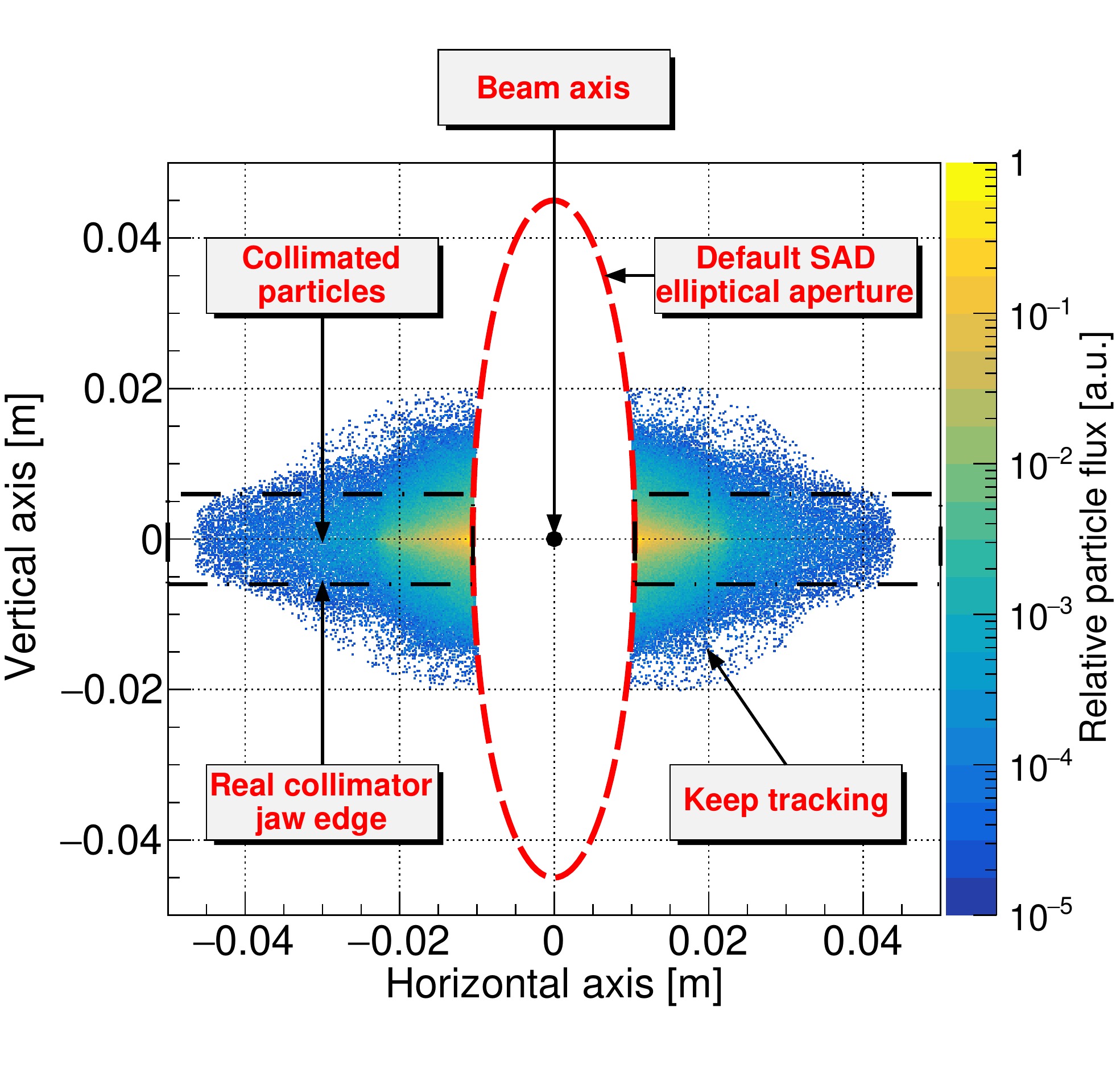}
\caption{\label{fig:fig10}Distribution of collimated beam particles at the LER D06H1 horizontal collimator. The red, dashed ellipse shows the original SAD collimator. The two black, dot-dashed rectangles show the newly implemented, more realistic collimator edge. The relative particle flux is shown in color. The bin size is $\SI{0.2}{mm}\times\SI{0.2}{mm}$.}

\end{figure}

Figure~\ref{fig:fig10} illustrates a particle distribution simulated with default (elliptical) and realistic (rectangular) profiles of the collimator. Particles that would previously hit the elliptical collimator and be classified as lost from the beam, can now pass the more realistic collimator and remain in the beam to be tracked by the simulation. Although the majority of the collimated beam is located at the tip of the collimator, particles far from the beam axis and outside the real edge of the collimator jaw can contribute to IR losses.

\subsection{\label{subsec:ParticleTipscattering}New collimator tip scattering simulation}

To implement beam particle scattering off of collimators jaws, referred to as collimator tip scattering, we carried out a dedicated Geant4 simulation and parameterized the results. Targets made of tungsten, tantalum, and copper were bombarded by \SIlist[list-units=single]{4;7}{GeV/c} positron and electron beams, respectively. The simulation results were then transformed into a matrix form and used in SAD with the new, more realistic collimator profiles described above.

\begin{figure}[htbp]
  \subfloat{%
  \begin{tabular}{c}
    \includegraphics[width=\linewidth]{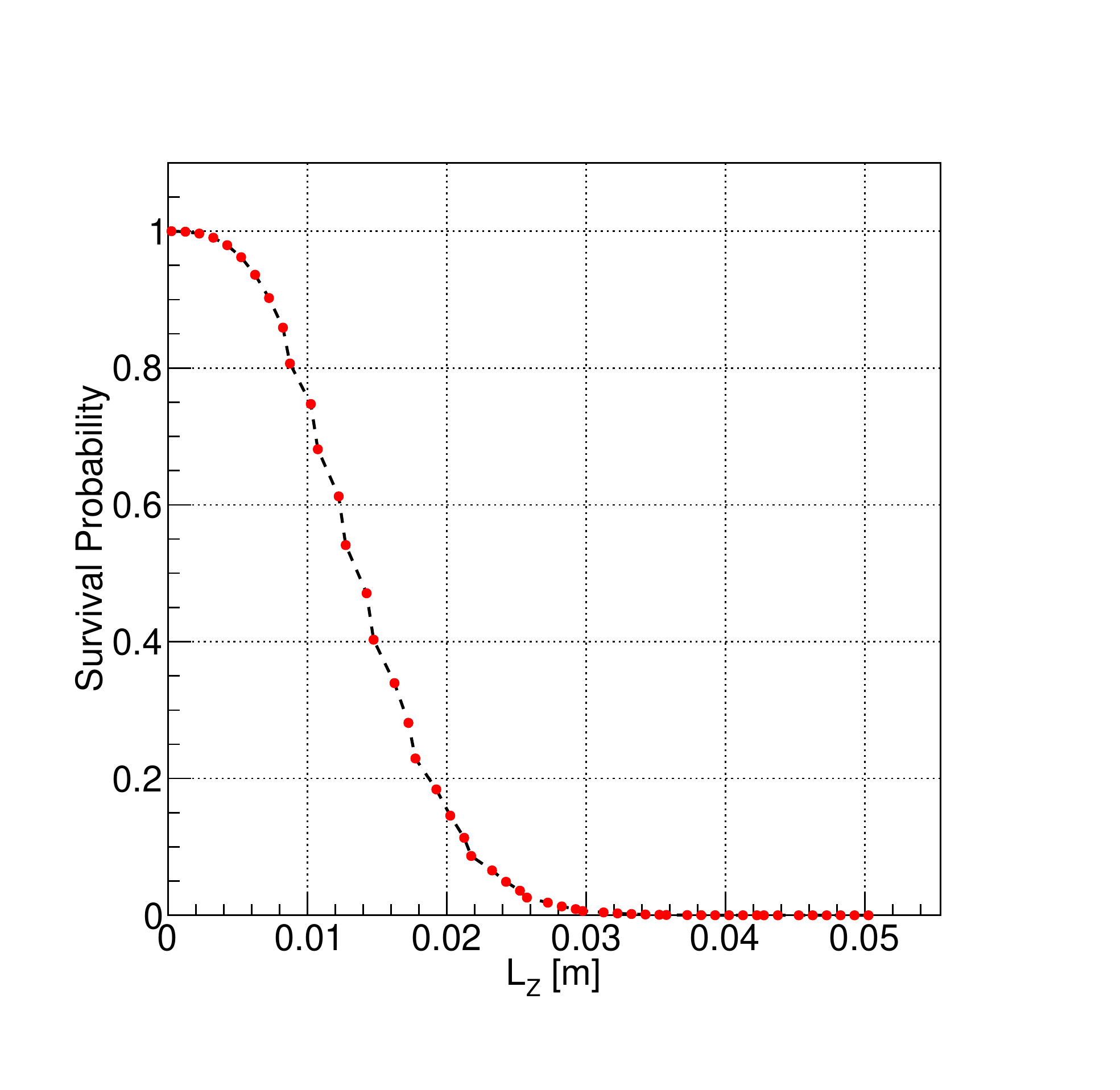}\\
    (a)~Survival probability\label{fig:fig11a}\\
  \end{tabular}
  }
  \\
  \subfloat{%
  \begin{tabular}{c}
    \includegraphics[width=\linewidth]{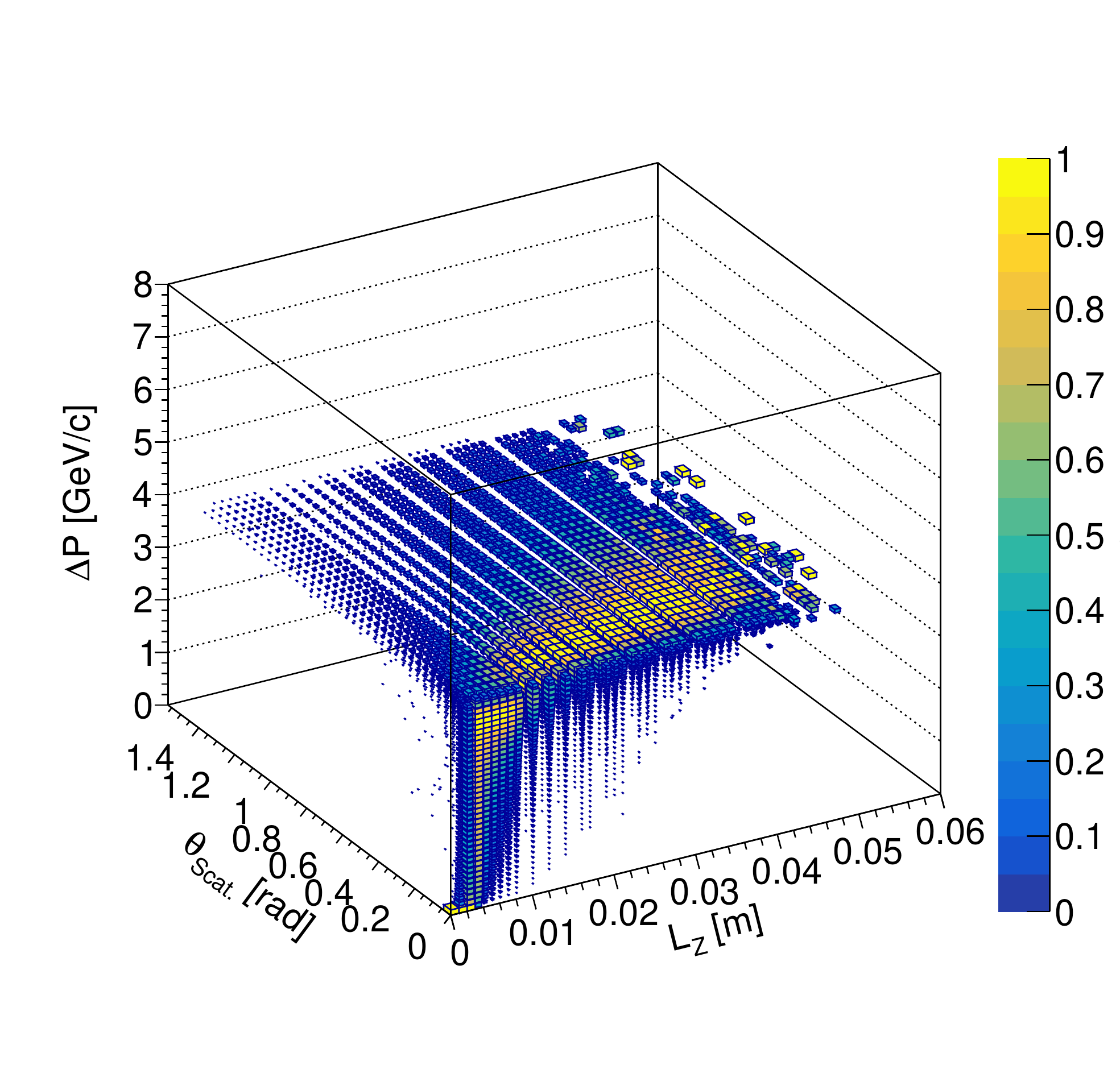}\\
    (b)~Momentum change\label{fig:fig11b}\\
  \end{tabular}
  }
  \caption{\label{fig:fig11}Geant4 simulation results for \SI{4}{GeV/c} positrons interacting with tungsten. (a) Particle survival probability versus path length inside the jaw, $L_{\rm Z}$. Statistical error $< 1\%$. (b) Momentum change, $\Delta P$, versus scattering angle, $\theta_{\rm Scat.}$, and path length inside the jaw. Each slice of $L_{\rm Z}$ is normalized so that its maximum value in the $\Delta P $-$\theta_{\rm Scat.}$ plane is unity. Therefore, the color and the size of each box (bin) represent the relative probability for a scattered particle with a given given $L_{\rm Z}$ to obtain a particular $\Delta P$ and $\theta_{\rm Scat.}$. Bin size is $\SI{1}{mm} \times \SI{40}{mrad} \times \SI{100}{MeV/c}$.}
    
\end{figure}

Due to the tapered collimator structure, particles intercepted by the jaw at different radial distances from the beam center traverse different path lengths, $L_{\rm Z}$, inside the head along the direction of the beam. Therefore, the survival probability as a function of  $L_{\rm Z}$ was included in SAD as well (Figure~\ref{fig:fig11a}). The induced angular kick and momentum change in the case of scattering are shown in Figure~\ref{fig:fig11b}. When a simulated beam particle collides with a collimator jaw, we apply a change to the particle's momentum ($\Delta p_{\rm X}/p_{\rm 0}$, $\Delta p_{\rm Y}/p_{\rm 0}$ and $\Delta p/p_{\rm 0}$), which depends on the path length inside the jaw. The momentum change is drawn from a parameterized probability density function which is different for each slice of $L_{\rm Z}$, and has a maximum equal to one (Fig.~\ref{fig:fig11b}). Afterwards, the particle is tracked again through the machine lattice until it is lost or the simulation reaches the maximum number of turns. Due to the Mathematica-like nature of SAD, which facilitates matrix calculations, our enhanced algorithm does not significantly increase the computing time for typical simulation statistics.

The realistic collimator shape and tip scattering were implemented only for \textit{SuperKEKB-type} collimators, while \textit{KEKB-type} collimators are modelled as ideal absorbers of particles outside the default elliptical aperture.

\section{\label{sec:ReportOnValidationAndAchievements}Improved predictions and experimental validation}

In this section, we review the key results of the improved particle tracking and collimator simulations. The implementation of more realistic collimator shapes solves a longstanding problem with the predicted background rates. In addition, we introduce a new method for the collimation system optimization, based on the new sequential tracking algorithm, which significantly reduces computing time and helps operators set collimator positions for efficient background reduction. Finally, we experimentally validate the improved collimator simulation by comparing simulated backgrounds against measured IR dose rates during a collimator aperture scan. We find good agreement between experimental data and Monte-Carlo simulation results.

\subsection{\label{subsec:ResolvedAlongstandingProblem}Resolving a longstanding problem}

The improved particle tracking and collimator shape implementation significantly improve the agreement between experimental data and Monte-Carlo simulation. Whereas previously discrepancies as large as two or three orders of magnitude were observed~\cite{REF3}, these have now been reduced to a factor of ten or less.

\begin{table*}[htbp]
\caption{\label{tab:tab4}Comparison of the predicted SuperKEKB beam particle IR loss rates ($\mathcal{R}$) and beam lifetime ($\tau$) from SAD simulations with \textit{elliptical} (\textit{ellipt.}) and \textit{realistic} (\textit{real.}) collimator profiles. Simulated collimator apertures and optics were those of May 2020: LER $\beta^{*}_{\rm H/V} = \SI{80/1}{mm}$, HER $\beta^{*}_{\rm H/V} = \SI{60/1}{mm}$ with the Crab Waist scheme.}
\begin{ruledtabular}
\begin{tabular}{ cccccccc }
Background 	& \multirow{2}{*}{Parameters} 	& HER 		& HER 	& HER ratio & LER 		& LER & LER ratio 	\\
source			& 						& \textit{ellipt.}	& \textit{real.} & \textit{real./ellipt.}	& \textit{ellipt.} 	& \textit{real.} & \textit{real./ellipt.} \\
\hline
\multirow{2}{*}{Coulomb}
& $\mathcal{R}$ [MHz] & $0.23 \pm 0.01$ & $9.38 \pm 0.30$ & $40.78 \pm 2.20$ & $55.25 \pm 0.99$ & $57.48 \pm 1.03$ & $1.04 \pm 0.02$\\
& $\tau$ [min] & $261.00 \pm 20.26$ & $247.81 \pm 18.67$ & $0.95 \pm 0.10$ & $28.61 \pm 1.71$ & $28.34 \pm1.67$ & $0.99 \pm 0.08$\\
\hline
\multirow{2}{*}{Bremsstrahlung}
& $\mathcal{R}$ [MHz] & $0.52 \pm 0.01$ & $0.55 \pm 0.01$ & $1.06 \pm 0.03$ & $4.74 \pm 0.06$ & $4.66 \pm 0.06$ & $0.98 \pm 0.02$\\
& $\tau$ [min] & $7513.63 \pm 86.29$ & $7532.44 \pm 92.27$ & $1.00 \pm 0.02$ & $1524.39 \pm 8.93$ & $1523.32 \pm 8.60$ & $1.00 \pm 0.01$\\
\hline
\multirow{2}{*}{Touschek}
& $\mathcal{R}$ [MHz] & $21.46 \pm 0.74$ & $48.06 \pm 8.93$ & $2.24 \pm 0.42$ & $26.80 \pm 2.33$ & $33.93 \pm 2.77$ & $1.27 \pm 0.15$\\
& $\tau$ [min] & $59.48 \pm 0.46$ & $58.65 \pm 0.46$ & $0.99 \pm 0.01$ & $6.75 \pm 0.04$ & $6.75 \pm 0.04$ & $1.00 \pm 0.01$\\
\end{tabular}
\end{ruledtabular}
\end{table*}

Table~\ref{tab:tab4} lists the rates of beam particles lost at the IR, defined as $|S| < \SI{4}{m}$, for two different profiles of the simulated collimators in SAD. We see that switching from the default (\textit{elliptical}) to the improved (\textit{realistic}) profile, Touschek and beam-gas IR loss rates increase by factors up to 40, while the beam lifetimes are unchanged. Note that measured IR backgrounds are proportional to the IR loss rate, but also depend on the detailed IR loss distribution. The increased IR loss rate is mainly due to the beam particles, which were previously stopped in simulation by the idealized, elliptical collimators, but which in reality are able to pass outside the rectangular collimator jaw. This component is identified with the "keep tracking" label in Figure~\ref{fig:fig10}. 

Although only four out of twenty collimators in the HER are \textit{SuperKEKB-type}, placed in the D01 section, we find that the more realistic collimator shapes result in significantly larger HER beam losses; both the Coulomb and Touschek components increase. We explain the more substantial background increase in the HER compared to the LER as a net contribution from collimators narrower than the interaction region aperture. All horizontal collimators in both rings and three vertical collimators in the LER are narrower than the IR. Thus stray particles in the vertical plane not absorbed by the most upstream vertical LER collimator can be stopped by the other two downstream collimators before reaching the IR. In contrast, in the HER, only one vertical collimator, D01V1, is narrower than the IR aperture, while other HER vertical collimators in D09 and D12 sections are wider than the IR. Although Coulomb losses are distributed in both planes, Touschek losses mostly occur in the horizontal plane due to the non-zero horizontal dispersion and momentum deviation of scattered particles. However, after hundreds of machine turns, Touschek scattered particles that are not absorbed by horizontal collimators propagate from the horizontal to the vertical plane due to the coupling between horizontal and vertical betatron motions. Therefore, the change of the D01V1 collimator shape affects not only Coulomb but also Touschek beam losses. Also, D01V1 has a larger horizontal beta function (Tab.~\ref{tab:tab5}) than LER vertical collimators. Therefore, the modification of its shape in simulation enhances the rate of particles passing outside the rectangular collimator jaw in the horizontal plane, as there the beam is wider than the jaw's transverse width. The combination of all effects described above leads to a higher increase in HER backgrounds compared to the LER case.

\begin{table}[htbp]
\caption{\label{tab:tab5}SuperKEKB vertical collimator settings in May 2020, when the Crab Waist optics were used. The third and forth columns list the vertical (V) and horizontal (H) beta functions at the collimators, $\beta$. The fifth column shows the collimator aperture in units of the betatron beam size, $\sigma_{\rm \beta} = \sqrt{\beta\times\varepsilon}$, where $\varepsilon$ is the beam emittance. The smallest vertical apertures in the IR are $106\sigma_{\beta \rm V}$ and $55\sigma_{\rm \beta V}$ for LER and HER, respectively.}
\begin{ruledtabular}
\begin{tabular}{ ccccc }
Ring & Collimator & $\beta_{\rm V}$ [m] & $\beta_{\rm H}$ [m] & \textit{d} [$\sigma_{\rm \beta V}$] \\
\hline
\multirow{3}{*}{LER} & D06V1 & 67.35 & 14.63 & 58 \\
 & D06V2 & 20.57 & 9.96 & 82 \\
 & D02V1 & 13.95 & 10.89 & 75 \\
\hline
HER & D01V1 & 46.46 & 40.13 & 41 \\
\end{tabular}
\end{ruledtabular}
\end{table}

The improved collimator shapes in simulation thus resolved the longstanding deficit in predicted backgrounds described in Section~\ref{sec:Introduction}, which became apparent with the first SuperKEKB background measurements in 2016~\cite{REF3}. The most recent detailed results, including comprehensive data/MC calculations for all Belle~II sub-detectors for the 2019-2020 data-taking period will be reported in a future publication.

\subsection{\label{subsec:CollimationSystemOptimization}Collimation system optimization}

The collimation system optimization described in Section~\ref{subsec:OptimizationProcedure} involves the manual adjustment of each collimator aperture, making the procedure very time consuming. Due to many degrees of freedom present in the collimation system (10 and 20 collimators for LER and HER, respectively), performing such an optimization requires several hours of SuperKEKB operation without Belle~II data taking. Furthermore, non-optimal collimation system settings can occur due to local minima, and the phase advance collimation is only effective against scattering upstream of the collimator. Therefore, up to now, a dedicated Monte-Carlo code was used at KEK to run a separate simulation in SAD for each set of collimator potential apertures. Unfortunately, finding the optimal collimator configuration with this code takes several days. Our new sequential tracking framework, described above, now greatly reduces the computational time required for collimator optimization in simulation. The new procedure collects particle distributions at each collimator with fully open collimators, running the SAD simulation only once. Afterwards, an \textit{off-line} linear scan of collimators is performed using C/C++ and CERN-ROOT based scripts. The goal is to maximize the beam lifetime while maintaining the lowest possible IR losses. This new method allows us to find an optimal position for all jaws using just a few hours of CPU time. This new procedure was not possible with the original particle tracking implementation, and directly enabled by the sequential tracking framework.

\begin{figure}[htbp]
\centering
\includegraphics[width=\linewidth]{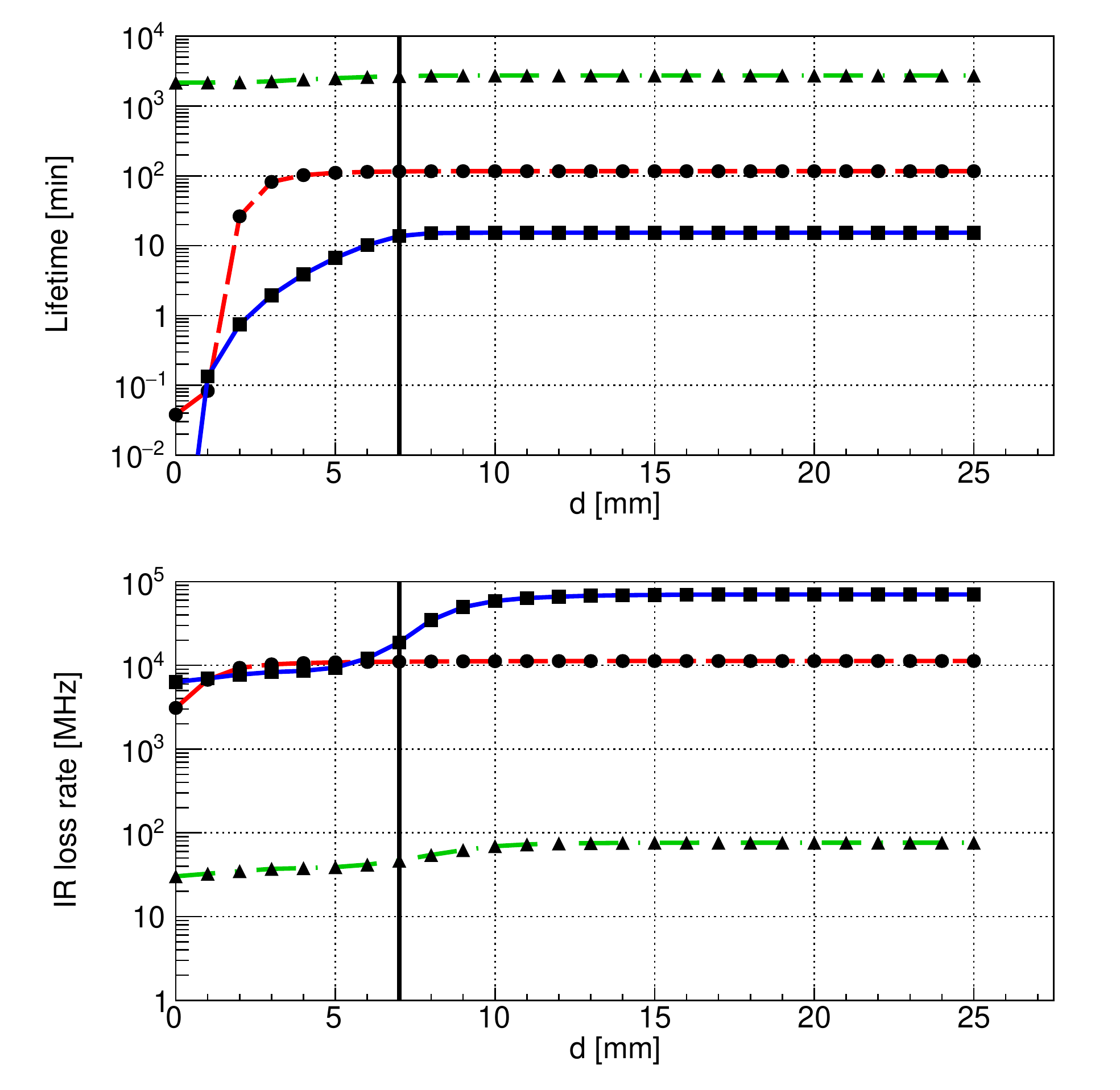}
\caption{\label{fig:fig6}Example of SAD simulation results for the optimization of an LER horizontal collimator, showing beam lifetime (top) and IR loss rate (bottom) versus  collimator aperture. The blue, solid line with rectangular markers shows Touschek scattering; the red, dashed line with circles shows Coulomb scattering; the green, dot-dashed line with triangles shows Bremsstrahlung scattering. The vertical black, solid line indicates the optimal aperture of the collimator.}

\end{figure}

Figure~\ref{fig:fig6} demonstrates the \textit{off-line} collimator scan procedure in simulation for an LER collimator. As the collimators are gradually closed, going from right to left in the figure, the collimator aperture at which the beam lifetime starts to decrease significantly is the optimal one. In this optimal configuration, the collimator mainly absorbs particles that potentially hit the IR aperture. Since these particles are going to be lost anyway, the beam lifetime is constant even if IR losses decrease for the aperture scan $d=\SI{10}{mm} \rightarrow \SI{7}{mm}$ in Figure~\ref{fig:fig6}. For a narrower aperture, the collimator stops particles that are not lost even after 1000 turns. Therefore, for the scan $d=\SI{7}{mm} \rightarrow \SI{2}{mm}$, we see beam lifetime degradation while the IR losses stay almost flat. Applying the same logic to all collimators, and iterating as needed, the optimal combined collimator configuration for each ring can be found.

\begin{figure}[htbp]
  \subfloat{%
  \begin{tabular}{c}
    \includegraphics[width=\linewidth]{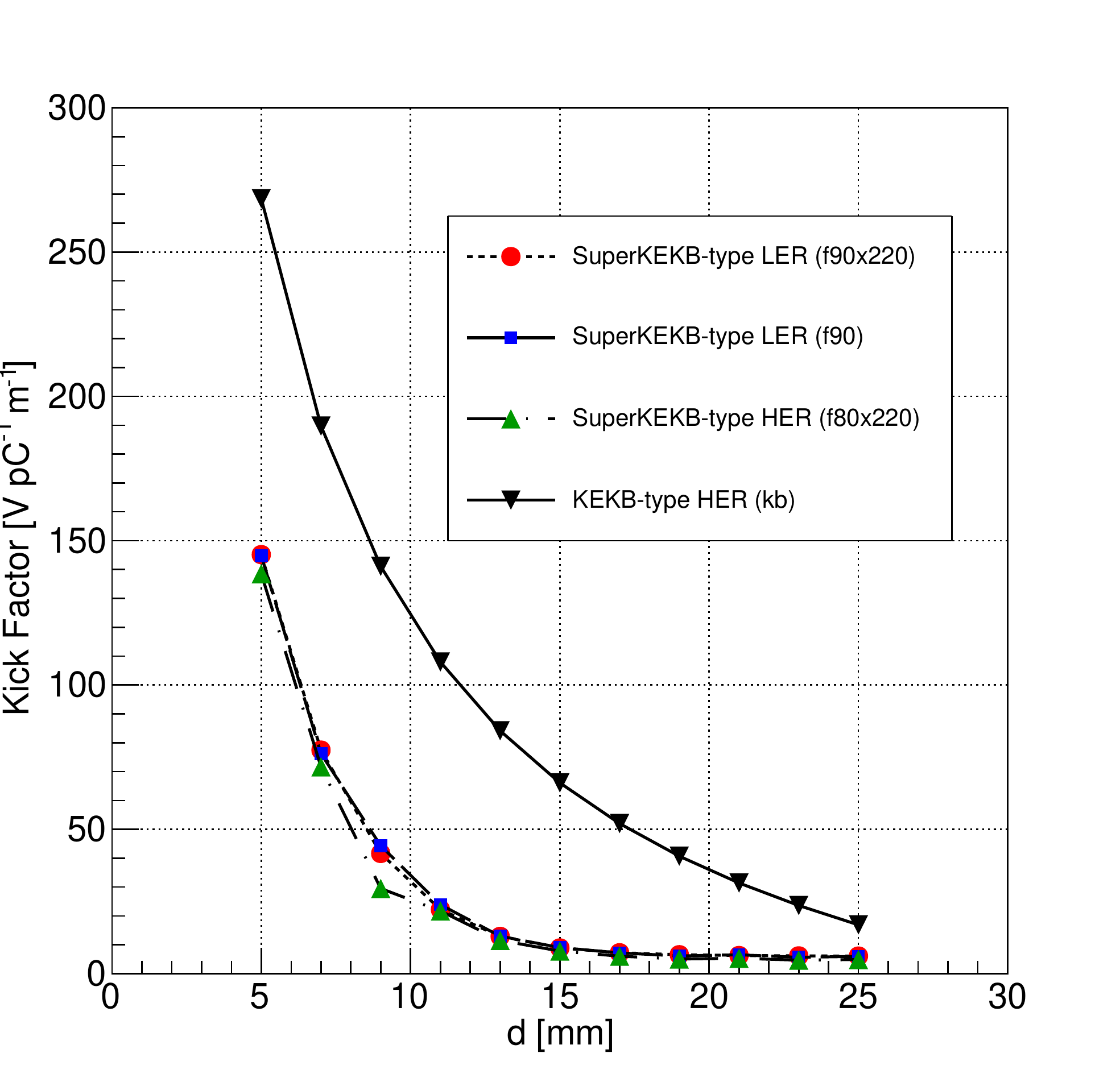}\\
    (a)~Horizontal collimators\label{fig:fig7a}\\
  \end{tabular}
  }
\\
  \subfloat{%
  \begin{tabular}{c}
    \includegraphics[width=\linewidth]{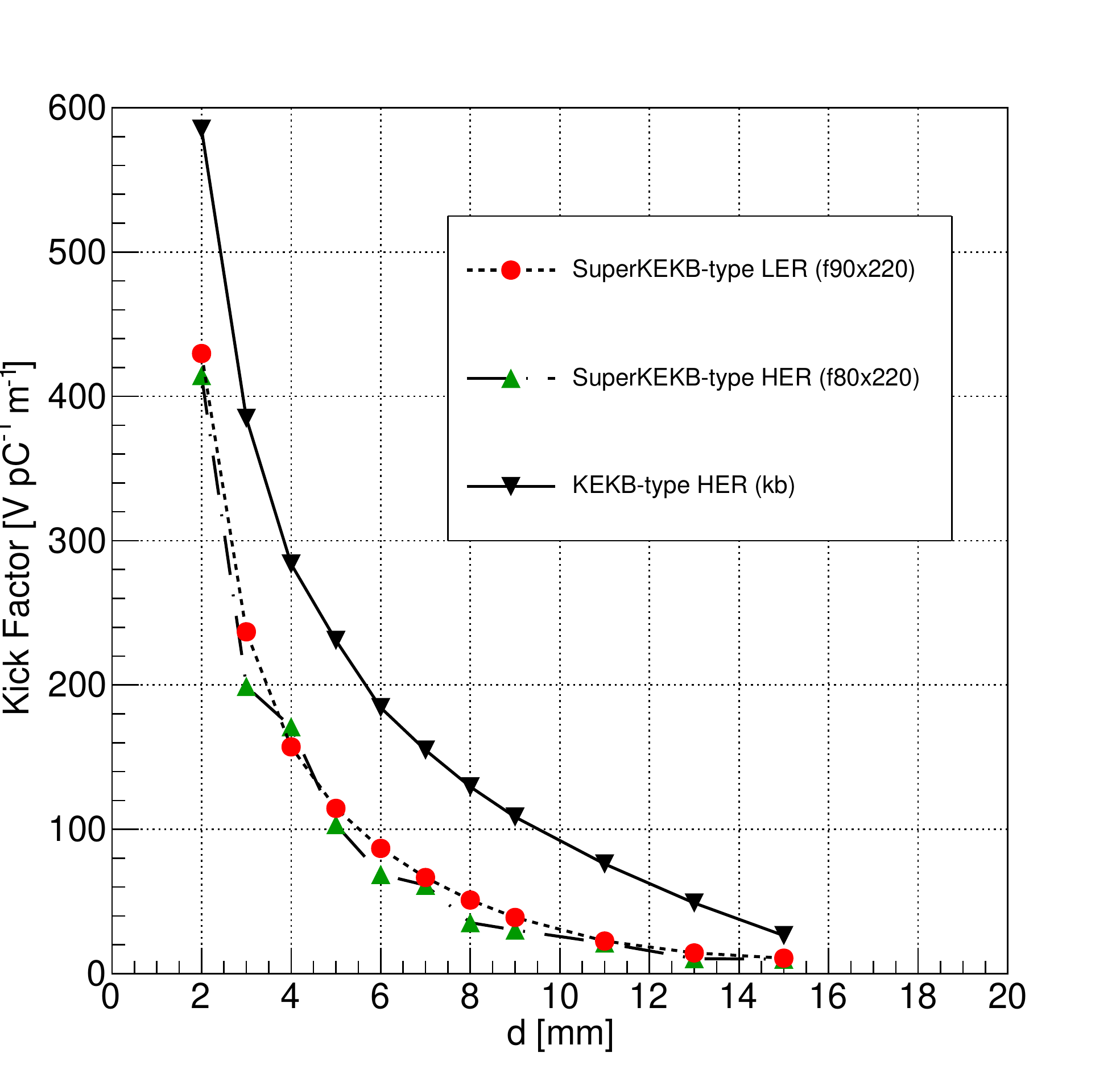}\\
    (b)~Vertical collimators\label{fig:fig7c}\\
  \end{tabular}
  }
  \caption{\label{fig:fig7}GdfidL simulation results for the horizontal (a) and vertical (b) collimator kick factor versus collimator aperture. The assumed bunch length is \SI{6}{mm}. The collimator cross-section is specified in brackets~\cite{REF1}.}
    
\end{figure}

Although minimum IR losses and a high, acceptable lifetime can be achieved in simulation by this procedure, the collimators also have to satisfy specific requirements to avoid what is known as transverse mode coupling instability (TMCI)~\cite{REF17}. TMCI is a result of the wake-field effect from bunches of charge travelling through the collimator aperture leading to the onset of the bunch current head-tail instability, which may enlarge the beam size. According to Ref.~\cite{REF15}, the bunch current must be limited, depending on the collimator aperture, in order to avoid TMCI:

\begin{equation}
I_{\rm thresh.} = \frac{8f_{\rm s}E/e}{\sum\limits_{j}\beta_{j}k_{j}(\sigma_{\rm S},d)},
\label{eq:eq1}
\end{equation}
where $I_{\rm thresh.}$ is the upper limit on the bunch current, $f_{\rm s}=\SI{2.13}{kHz}$ or $f_{\rm s}=\SI{2.80}{kHz}$ is the synchrotron frequency for LER or HER, respectively, $E$ is the beam energy, $e$ is the unit charge, $\beta_{j}$ and $k_{j}$ are the beta function and kick factor of the $j$-th collimator, respectively. The analytical formula for the kick factor as a function of the longitudinal beam size, $\sigma_{\rm S}$, and the collimator aperture, $d$, is given in Ref.~\cite{REF15}. Here, however, we account for the geometry of each collimator by using GdfidL~\cite{REF16} simulation results for the $k_{j}$ values, see Fig.~\ref{fig:fig7}.

\begin{figure}[htbp]
  \subfloat{%
  \begin{tabular}{c}
    \includegraphics[width=\linewidth]{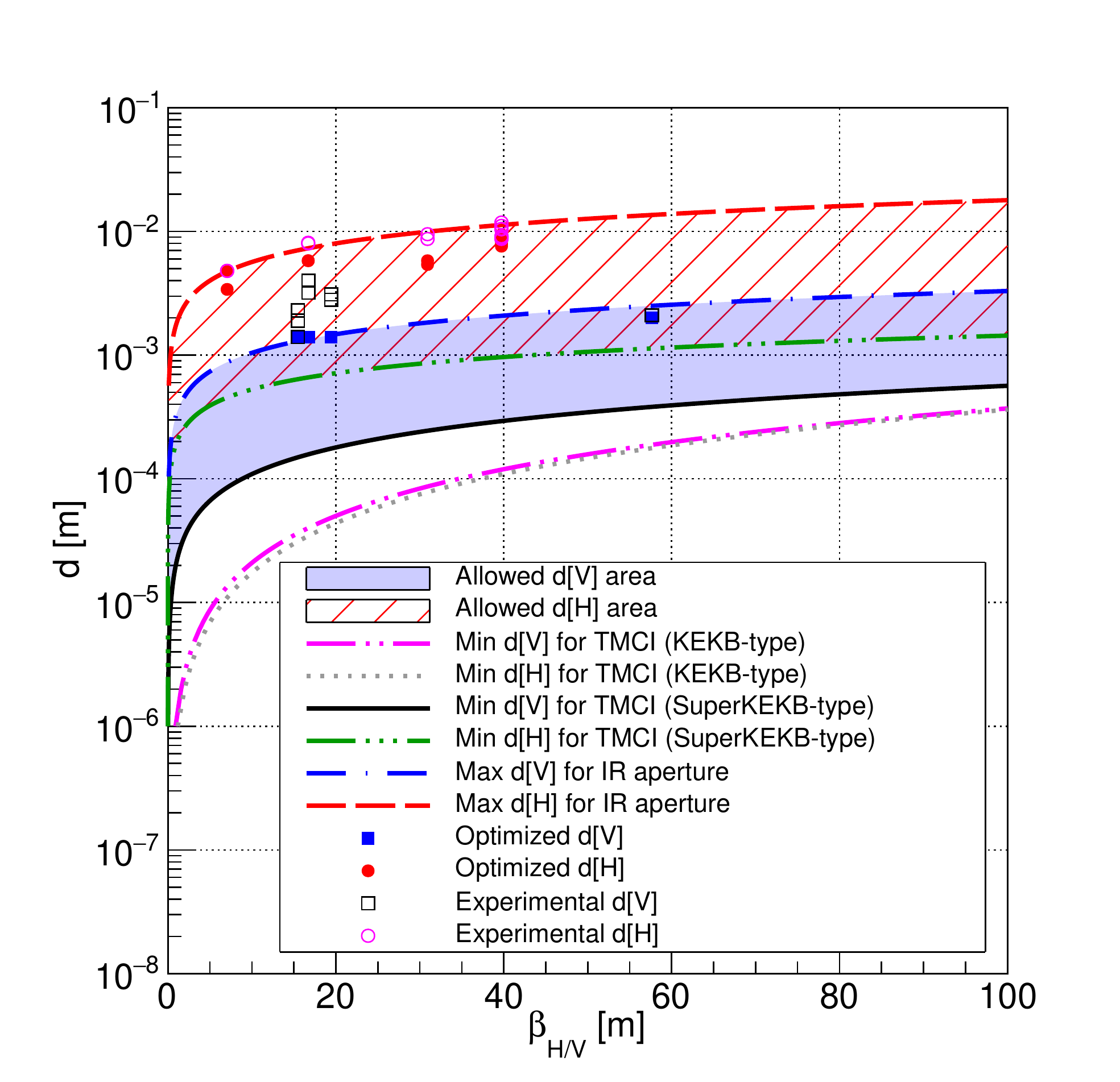}\\
    (a)~HER\label{fig:fig8a}\\
  \end{tabular}
  }
\\
  \subfloat{%
  \begin{tabular}{c}
    \includegraphics[width=\linewidth]{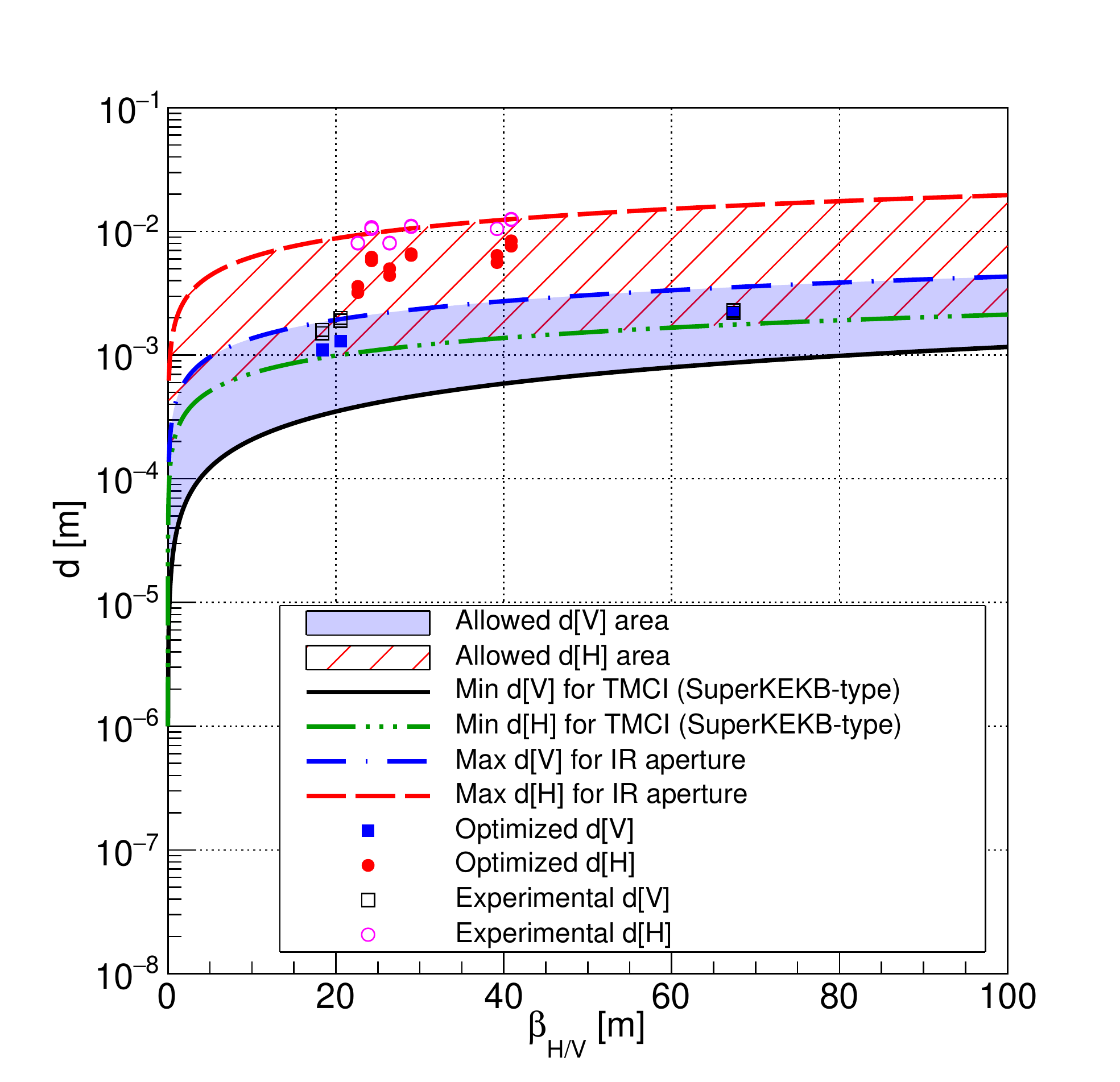}\\
    (b)~LER\label{fig:fig8b}\\
  \end{tabular}
  }
  \caption{\label{fig:fig8}SuperKEKB HER (a) and LER (b) collimator apertures and their constraints. Black, solid (green, triple-dot-dashed) and magenta, double-dot-dashed (gray, dotted) lines show the minimum allowed \textit{SuperKEKB-type} and \textit{KEKB-type} vertical (horizontal) collimator apertures at different beta function values, respectively, to avoid TMCI from a single collimator. The red, dashed and blue, dot-dashed lines show the maximum collimator aperture for horizontal and vertical collimators, respectively, beyond which we expect increased losses due to the IR aperture. Filled, red circles and filled, blue squares are optimized apertures of the horizontal and vertical collimators, respectively, based on simulation only; magenta, open circles and black, open squares are the experimental settings of the horizontal and vertical collimators used in June~2020.}
    
\end{figure}

Equation~\ref{eq:eq1} defines the highest acceptable bunch current ($I_{\rm b}$) and the narrowest possible collimator aperture before TMCI are expected. On the other hand, the aperture of the IR beam pipe dictates the maximum aperture of the collimator. Figure~\ref{fig:fig8} shows optimized apertures of LER and HER collimators, together with these two aperture constraints. To avoid TMCI and to protect the IR, horizontal (red circles) and vertical (blue squares) collimators have to be inside red, hatched and blue, filled areas, respectively. For a particular set of machine parameters, which are traditionally used in Belle~II for the beam background simulation, $I_{\rm LER/HER} = \SI{1.2/1.0}{A}$ and $N_{\rm bunch} = 1576$, the simulation-optimized collimator settings shown satisfy the bunch current requirements for both rings: (i) LER $I_{\rm b} = \SI{0.76}{mA} < I^{opt.}_{\rm thresh.} = \SI{1.15}{mA}$; (ii) HER $I_{\rm b} = \SI{0.63}{mA} < I^{opt.}_{\rm thresh.} = \SI{1.19}{mA}$. For SuperKEKB operation in June 2020, however, the experimental collimator settings used during operation are outside the allowed areas (Fig.~\ref{fig:fig8}, magenta opened circles and black opened squares), which implies they are ineffective for stray particle collimation. However, these settings permit the increase of the stored bunch current if the background level is acceptable. The maximum bunch current satisfying TMCI constraints would be $I^{\rm exp.}_{\rm thresh.} = \SI{1.52}{mA}$ in the LER and $I^{\rm exp.}_{\rm thresh.} = \SI{1.73}{mA}$ in the HER.

The optimized collimator configuration obtained from simulation can be considered an idealized target that would provide the lowest feasible IR backgrounds. However, the quality of the injected beam is one of the stumbling blocks, as it imposes additional restrictions on the range of jaw movement during the manipulation of the collimators. Understanding this limitation from first principles would be highly beneficial for realistic background planning, and will be pursued in the future.

Table~\ref{tab:tab2} compares the simulated IR loss rates for an experimental configuration of collimators (i.e., found by an operator) and an optimized collimator configuration obtained from simulation. We find that starting from the experimental configuration, an additional squeezing of collimators should reduce IR losses further. However, this is just a reference for operators since injection errors were not simulated. Therefore, even if the storage background can be significantly reduced by closing a collimator, unstable injection limits the aperture of collimators.

\begin{table*}[htbp]
\caption{\label{tab:tab2}Comparison of the SAD simulated IR losses ($\mathcal{R}$) and beam lifetime ($\tau$) for the experimental (\textit{exp.}) and optimized (\textit{opt.}) collimator settings. Beam optics were those of June 2020: $\beta^{*}_{\rm H/V} = \SI{60/0.8}{mm}$ with the Crab Waist scheme.}
\begin{ruledtabular}
\begin{tabular}{ cccccc }
Background 	& \multirow{2}{*}{Parameters} 	& HER 		& HER 	& LER 		& LER 	\\
source			& 						& \textit{exp.}	& \textit{opt.}	& \textit{exp.} 	& \textit{opt.} \\
\hline
\multirow{2}{*}{Coulomb}
& $\mathcal{R}$ [MHz] & $13.18 \pm 0.63$ & $3.71 \pm 0.23$ & $68.93 \pm 2.58$ & $13.66 \pm 0.74$ \\
& $\tau$ [min] & $281.54 \pm 31.15$ & $263.58 \pm 28.01$ & $38.33 \pm 3.08$ & $33.24 \pm 2.72$ \\
\hline
\multirow{2}{*}{Bremsstrahlung}
& $\mathcal{R}$ [MHz] & $0.43 \pm 0.01$ & $0.41 \pm 0.01$ & $3.19 \pm 0.08$ & $2.53 \pm 0.05$ \\
& $\tau$ [min] & $7921.77 \pm 122.92$ & $7819.97 \pm 122.22$ & $2326.91 \pm 21.33$ & $2191.14 \pm 24.68$ \\
\hline
\multirow{2}{*}{Touschek}
& $\mathcal{R}$ [MHz] & $32.36 \pm 9.46$ & $0.57 \pm 0.74$ & $67.42 \pm 3.81$ & $0.93 \pm 0.07$ \\
& $\tau$ [min] & $45.62 \pm 0.79$ & $41.34 \pm 0.51$ & $13.82 \pm 0.09$ & $8.59 \pm 0.06$ \\
\end{tabular}
\end{ruledtabular}
\end{table*}

\subsection{\label{subsec:ExperimentalValidation}Experimental validation}

In order to validate the improvements to the particle tracking software described above, we performed two consecutive SuperKEKB collimator aperture scans on June~27th, 2020. Two specific LER collimators were chosen for this study. D06V1 was chosen because it was the most recently installed vertical collimator at the time, intended to suppress the dominant IR background component from LER beam-gas scattering. D02H4 was chosen because it is the closest horizontal collimator to the IR, allowing us to check for IR contributions from tip scattering, and to validate the simulation of this background component. We scanned the aperture of each collimator in steps, measuring IR losses using diamonds detectors. The high voltage was kept off in all Belle~II sub-detectors to avoid damage to sensitive electronics during collimator manipulation. Table~\ref{tab:tab3} summarizes the machine settings, where the narrowest IR element is QC1 at $S \approx \SI{1}{m}$ from the IP. QC1 is a superconducting quadrupole magnet of the final focusing system that has an aperture of 34$\sigma_{\rm \beta H}$ and 92$\sigma_{\rm \beta V}$ in the horizontal and vertical planes, respectively, where $\sigma_{\rm \beta} = \sqrt{\beta\times\varepsilon}$ stands for the betatron beam size.

\begin{table*}[htbp]
\caption{\label{tab:tab3}SuperKEKB settings for the LER collimator studies conducted on June~27, 2020 with the Crab Waist scheme. Indexes $H$ and $V$ indicate horizontal and vertical planes, respectively. $\beta$ is the beta function; $\alpha = -0.5 \times d\beta/dS$ is the correlation function; $D$ is the dispersion function; $D'$ is the chromatic derivative of $D$; $\Delta\mu$ is the phase advance; $\varepsilon$ is an emittance; $I$ is the beam current; $N_{\rm b}$ is the number of bunches; Inj. rep. is the injection repetition rate; $Q$ is the betatron tune; $\sigma_{\rm S}$ is the bunch length; \textit{P} is the particle momentum.}
\begin{ruledtabular}
\begin{tabular}{ ccccccc }
Element  & $S$ [m] & $\beta_{\rm H/V}$ [m] & $\alpha_{\rm H/V}$ & $D_{\rm H/V}$ [m] & $D_{\rm H/V}'$&$\Delta\mu_{\rm H/V}$ [$1/2\pi$]\\
\hline
QC1 & 0 & 28.56/978.06 & -113.30/-20.80 & -2.60e-5/-2.70e-4 & 1.34e-4/6.10e-4 & 0/0 \\
D02H4 & 15 & 26.40/16.50 & -0.54/-11.62 &  -4.43e-1/-1.72e-3 &  8.41e-2/-2.50e-4 & 0.04/0.98 \\
D06V1 & 1147 & 14.64/67.35 & -3.42/20.26  & 5.16e-1/-7.20e-9 & -7.28e-2/-1.10e-9 & 0.81/0.98\\
\toprule
$\varepsilon_{\rm H/V}$ $[\rm pm \cdot rad]$ & $I$ [mA] & $N_{\rm b}$ & Inj. rep. [Hz] & $Q_{\rm H/V}$ & $\sigma_{\rm S}$ [mm] & \textit{P} [GeV/c]\\
\hline
3300/22 & 200 & 978 & 12.5 & 45.525/43.581 & 5.0 & 4\\
\end{tabular}
\end{ruledtabular}
\end{table*}

\subsubsection{\label{subsubsec:VerticalCollimatorScan}Vertical collimator scan}

D06V1 was installed in January 2020 with a tungsten head and was upgraded with a low atomic number (low-\textit{Z}) material (graphite) head in September 2020. Since this upgrade occurred after the June~27th 2020 study described above, it is possible to compare both measured and simulated outcomes for a given collimator before and after this upgrade, making D06V1 an ideal candidate for particle tracking code verification.

\begin{figure*}[htbp]
  \subfloat{%
  \begin{tabular}{>{\centering}p{0.48\linewidth}}
    \includegraphics[width=\linewidth]{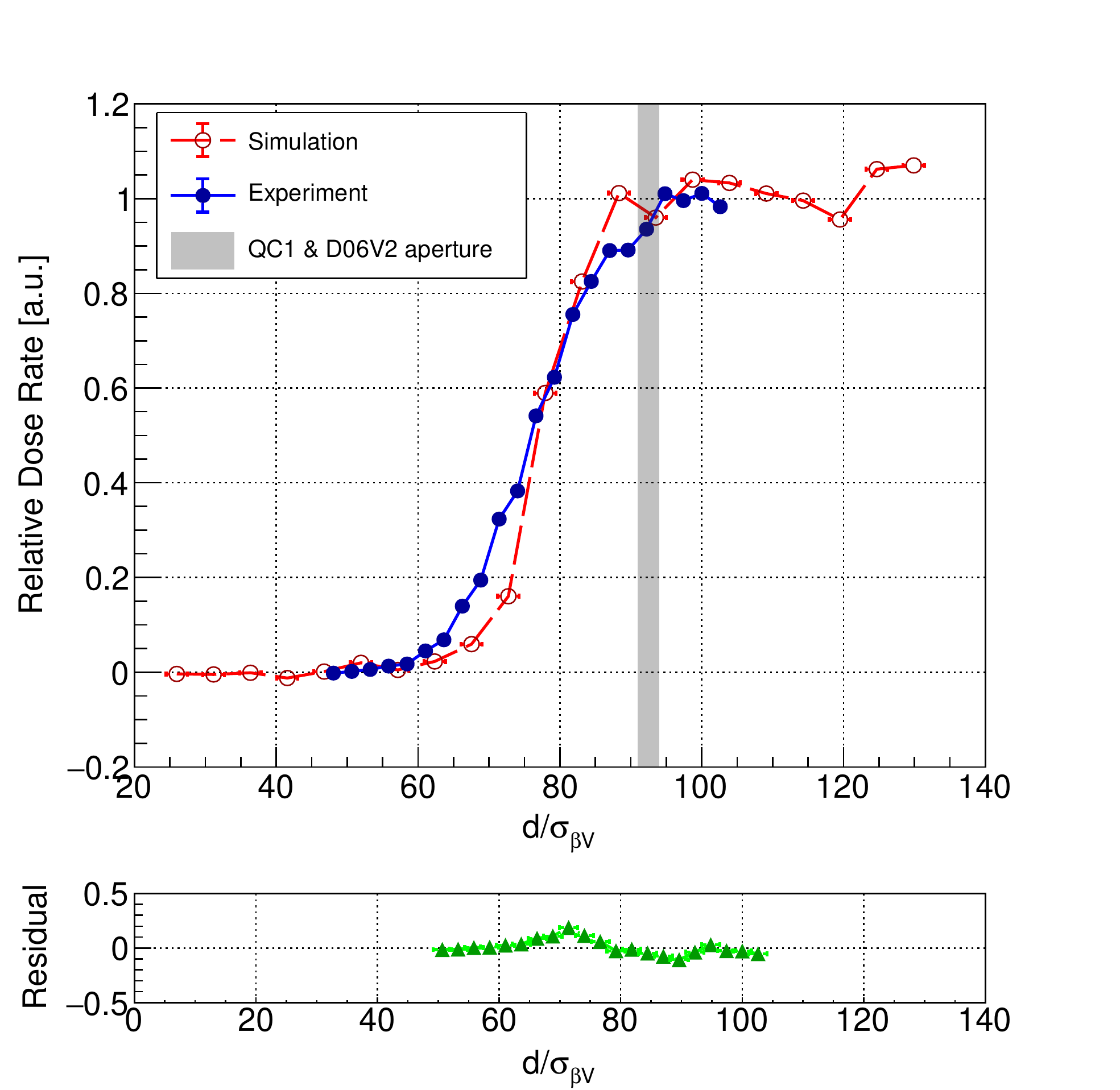}\\
    (a)~QCS-FW detectors (QCS-FW-135 \& QCS-FW-225)\label{fig:fig16a}\\
  \end{tabular}
  }
  \hfill
  \subfloat{%
  \begin{tabular}{>{\centering}p{0.48\linewidth}}
    \includegraphics[width=\linewidth]{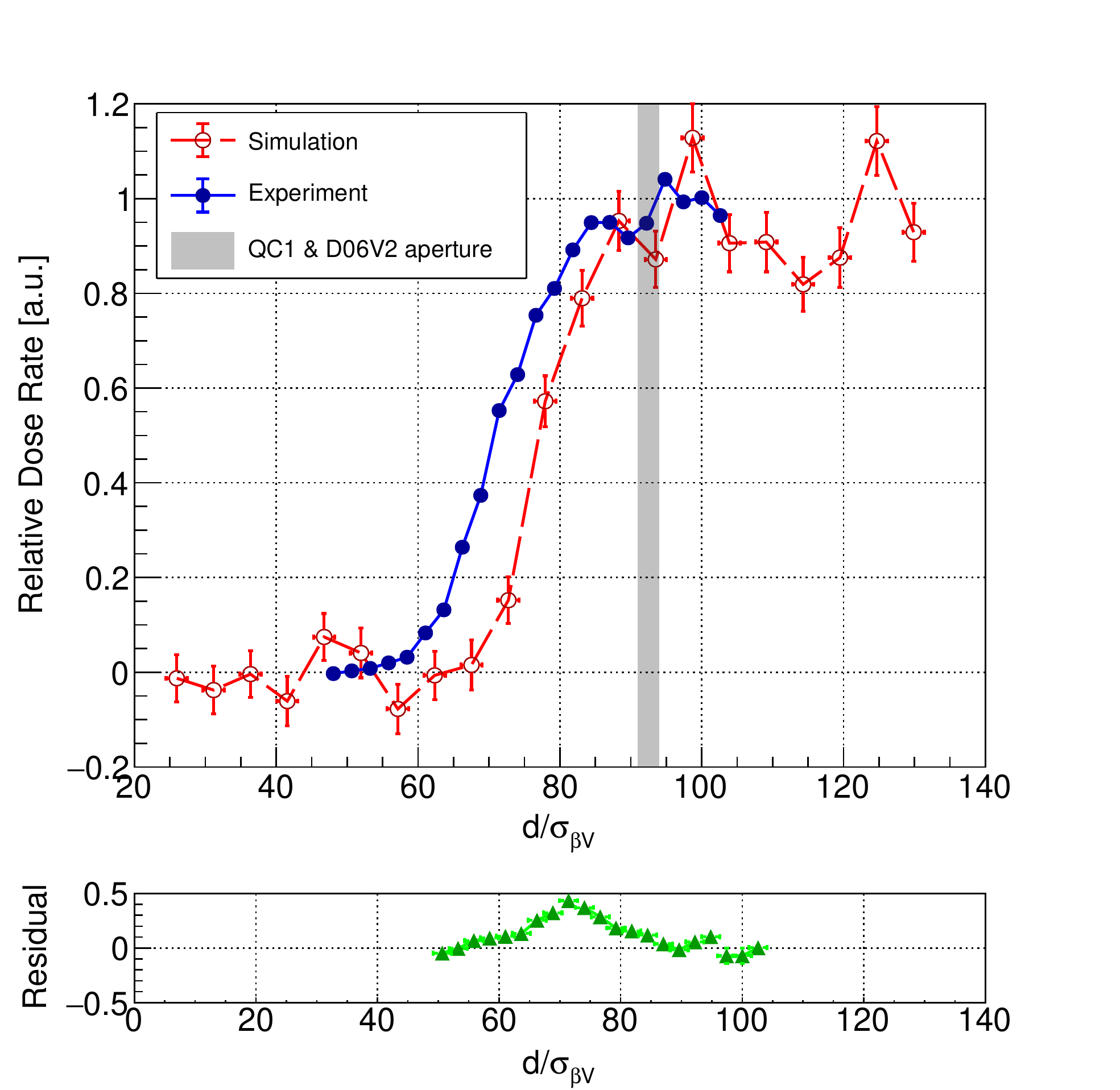}\\
    (b)~BP-FW/BW detectors (BP-BW-215, BP-BW-325 \& BP-FW-215)\label{fig:fig16b}\\
  \end{tabular}
  }
  \caption{\label{fig:fig16}Relative total dose rate for two groups of diamond detectors in the interaction region versus D06V1 collimator aperture, $d=(|d_{1}| + |d_{2}|)/2$, in units of $\sigma_{\rm \beta V}$. Top: the red, dashed line shows simulation results; the blue, solid line is experimental data; the grey, vertical band is the aperture of the QC1 magnet and D06V2 collimator. Bottom: Residual dose rate (experimental minus simulation).}
    
\end{figure*}

Figure~\ref{fig:fig16} shows the measured IR background versus  D06V1 aperture in units of $\sigma_{\rm \beta V}$, where we assume that the aperture is symmetric, i.e. $|d_{1}| = |d_{2}|$. Two specific sets of diamond detectors, QCS-FW and BP-FW/BW, were chosen for the background monitoring since they are the most sensitive to the LER background in the interaction region. The name of each diamond detector indicates its location: QCS is the superconducting magnet of the final focusing system; FW (BW) denotes the forward (backward) side of the Belle~II detector; 135, 215 or 225 are azimuth angle positions in degrees, see Fig.~\ref{fig:fig9}. To compare simulation and experimental data accurately, we normalize the two distributions. First, we define the average value in the \textit{valley} region, $d < 52\sigma_{\rm \beta V}$, as a pedestal, which we subtract from all data points. This ensures both pedestals are equal to zero. Then each data point is divided by the average value of the \textit{plateaus} at $91\sigma_{\rm \beta V} < d < 104\sigma_{\rm \beta V}$, which makes the two \textit{plateau} equal to one.

With a sufficiently small aperture, scattered particles that could have hit the IR are collimated (i.e., absorbed or scattered), and hence the background remains roughly constant. At about $60\sigma_{\rm \beta V}$, IR backgrounds start to increase as D06V1 intercepts a smaller fraction of the stray beam particles. The resulting observed dose rate versus aperture is close to \textit{S}-shaped because of the Gaussian-like beam profile. At this point, the collimator still represents the primary obstacle on the beamline since it has the narrowest aperture. A wider opening increases the number of particles that hit the IR vacuum pipe. At about $85\sigma_{\rm \beta V}$,  background losses begin to saturate because particles that were previously collimated by D06V1 are now redistributed and lost at other apertures after several turns in the machine. Around $90\sigma_{\rm \beta V}$, however, there is a noticeable $\sim 10\%$ increase in backgrounds, which does not follow the \textit{S}-shaped trend. This is caused both by QC1, which has an aperture of $92\sigma_{\rm \beta V}$, and by the D06V2 collimator with an aperture of $91-94\sigma_{\rm \beta V}$ ($\Delta\mu_{\rm V} = 0.34~[1/2\pi]$). D06V2 is \SI{20}{m} downstream of D06V1 (Fig.~\ref{fig:fig2}). The range  $91-94\sigma_{\rm \beta V}$, which includes the QC1 and D06V2 apertures, is highlighted in grey in Figure~\ref{fig:fig16}.

\begin{figure*}[htbp]
  \subfloat{%
  \begin{tabular}{c}
    \includegraphics[width=0.48\linewidth]{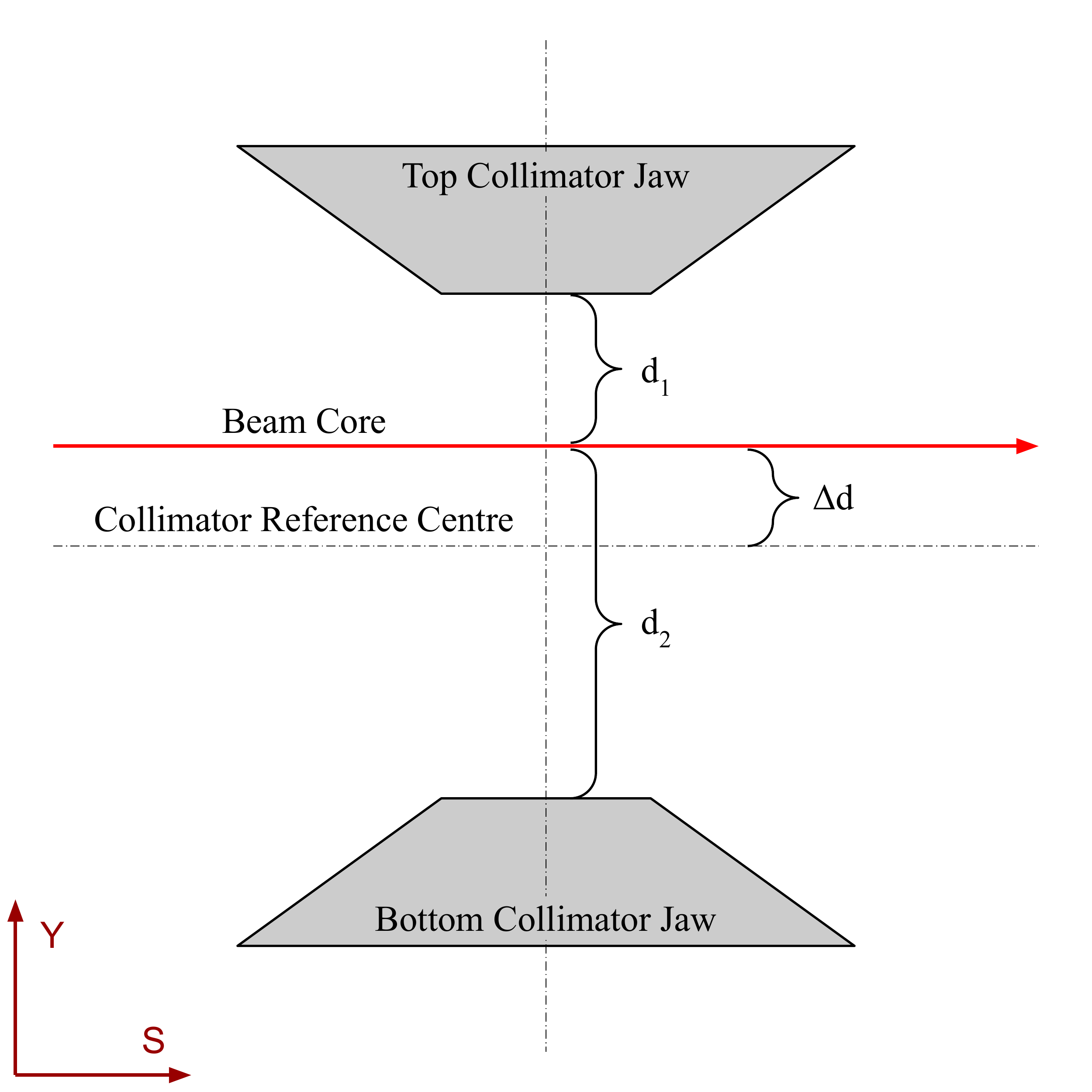}\\
    (a)~Collimator misalignment\label{fig:fig18a}\\
  \end{tabular}
  }
  \hfill
  \subfloat{%
  \begin{tabular}{c}
    \includegraphics[width=0.48\linewidth]{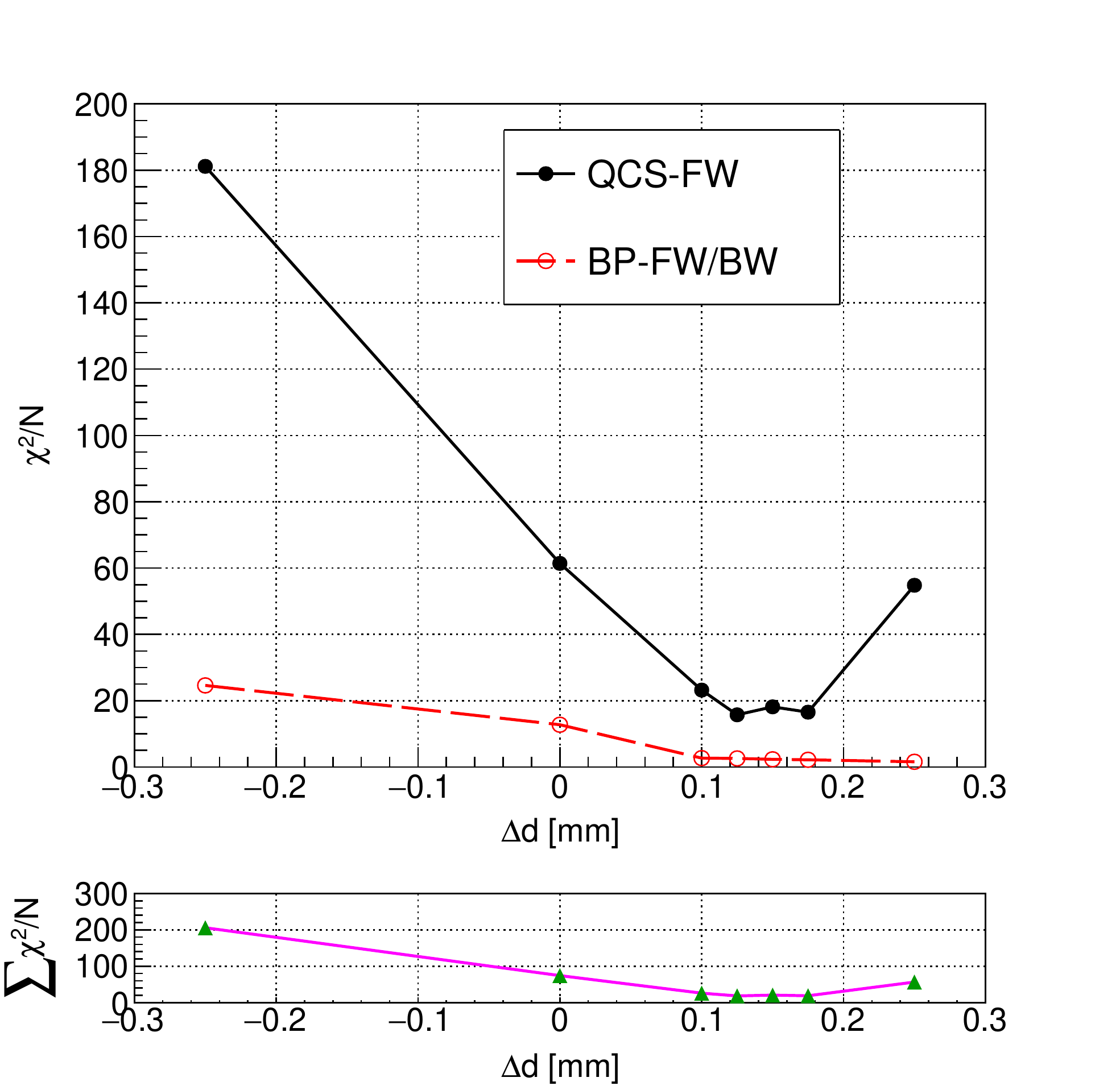}\\
    (b)~Simulation goodness\label{fig:fig18b}\\
  \end{tabular}
  }
  \caption{\label{fig:fig18}Vertical misalignment of the D06V1 collimator and its effect. (a) Illustration of the vertical offset ($\Delta d$) between the position reference of the collimator and the beam core. (b) The goodness of the simulation versus vertical offset. Top: reduced chi-square for two sets of diamond detectors; bottom: sum of those two reduced chi-squares versus vertical offset.}
\end{figure*}

The discrepancy between experimental and simulated data in Fig.~\ref{fig:fig16} may be due to the limited vertical \textit{SuperKEKB-type} collimator alignment accuracy of \SI{\pm 0.5}{mm}. Simulation shows that a vertical offset ($\Delta d$, Fig.~\ref{fig:fig18a}) between the coordinate reference of D06V1 and the beam center (defined by the beam position monitors of the nearest quadrupole magnets) can cause exactly such differences, specifically in the form of a modified slope of the measured \textit{S}-shape. To find the true vertical offset, a set of MC simulations was performed, where we apply different values of $\Delta d$ to the collimator aperture in SAD. We define the goodness of the simulation for each of the two sets of diamond detectors as:

\begin{equation}
\chi^{2}/N = \frac{1}{N}\sum_{i=1}^{N}\left( \frac{Res.[i]}{\Delta Res.[i]} \right)^{2},
\label{eq:eq7}
\end{equation}
where $Res.[i]$ and  $\Delta Res.[i]$ are the dose rate residual and its statistical uncertainty at a given aperture $d[i]$, while $N$ is the number of data points where $Res.$ was calculated (Fig.~\ref{fig:fig16}, bottom). Fig.~\ref{fig:fig18b} shows the goodness of the simulation versus vertical offset.
 
\begin{figure*}[h!tbp]
  \subfloat{%
  \begin{tabular}{c}
    \includegraphics[width=0.48\linewidth]{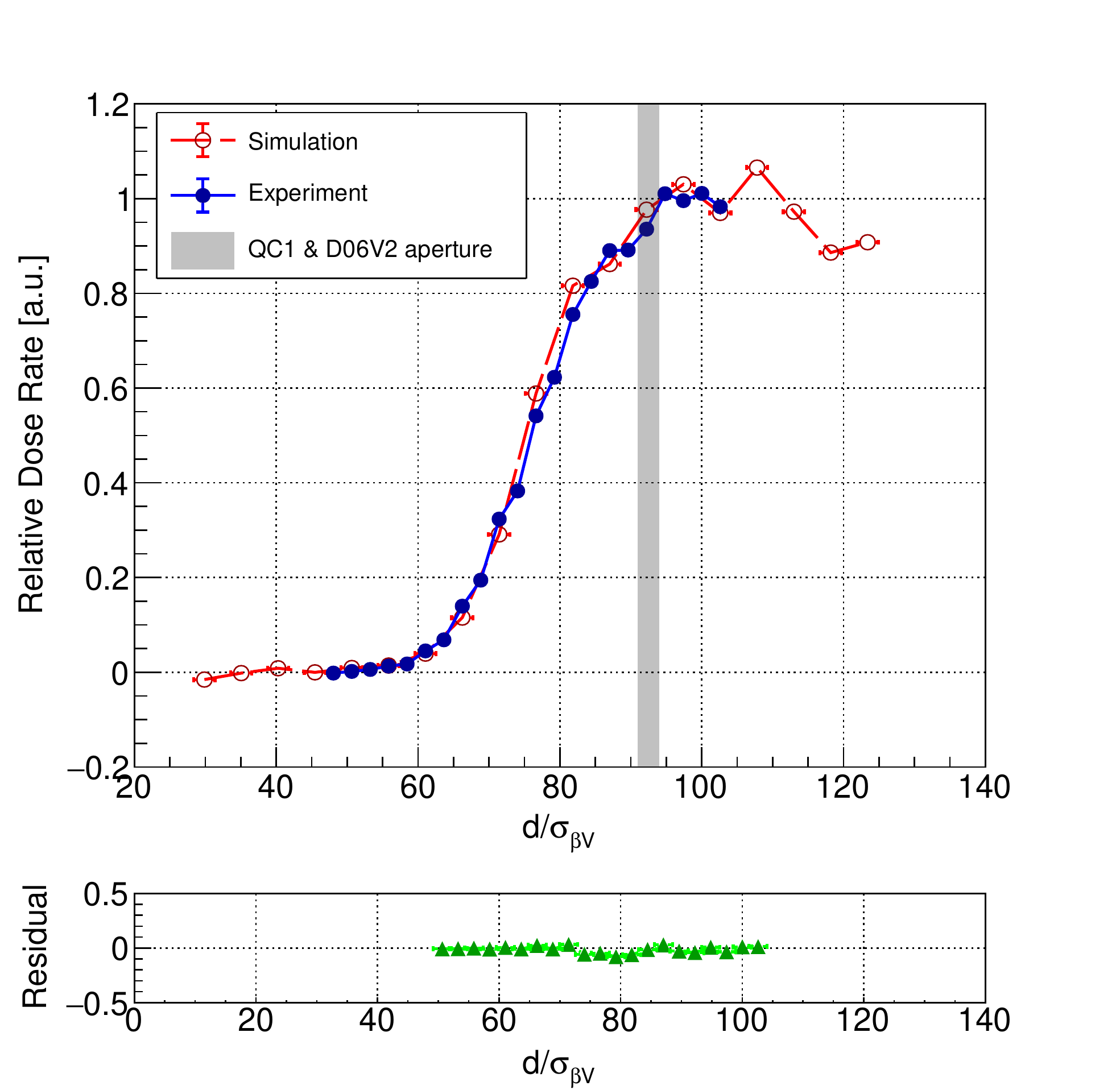}\\
    (a)~QCS-FW diamond detectors\label{fig:fig19a}\\
  \end{tabular}
  }
  \hfill
  \subfloat{%
  \begin{tabular}{c}
    \includegraphics[width=0.48\linewidth]{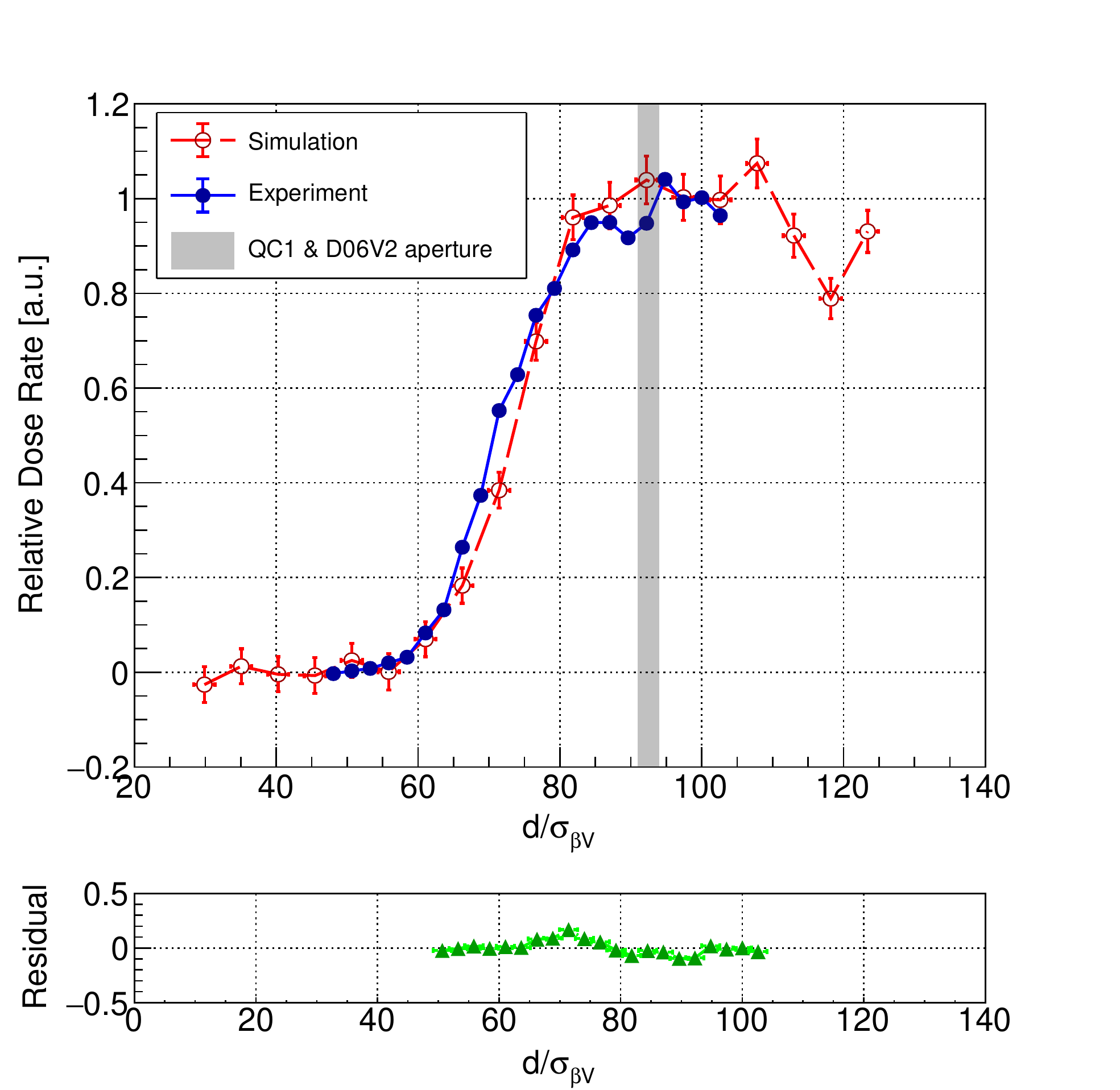}\\
    (b)~BP-FW/BW diamond detectors\label{fig:fig19b}\\
  \end{tabular}
  }
  \caption{\label{fig:fig19}Relative total dose rate in the interaction region versus D06V1 collimator aperture with a vertical offset $\rm\Delta d = \SI{0.15}{mm}$.}
    
\end{figure*}

The simulated collimator offset where $\sum\chi^{2}/N$ reaches a minimum suggests the optimal setting for the most accurate simulation of the experiment. Figure~\ref{fig:fig18b} shows that the optimal vertical offset is $\Delta d_{\rm opt.} = \SI[separate-uncertainty]{150 \pm 25}{\upmu m}$, which means that the top jaw is closer to the beam core than the bottom jaw. Figure~\ref{fig:fig19} shows the simulated collimator scan when this offset value is used to simulate the D06V1 collimator, i.e. so that $|d_{1}| + 2\Delta d_{\rm opt.} = |d_{2}|$ in the simulation.

\begin{figure}[htbp]
\centering
\includegraphics[width=\linewidth]{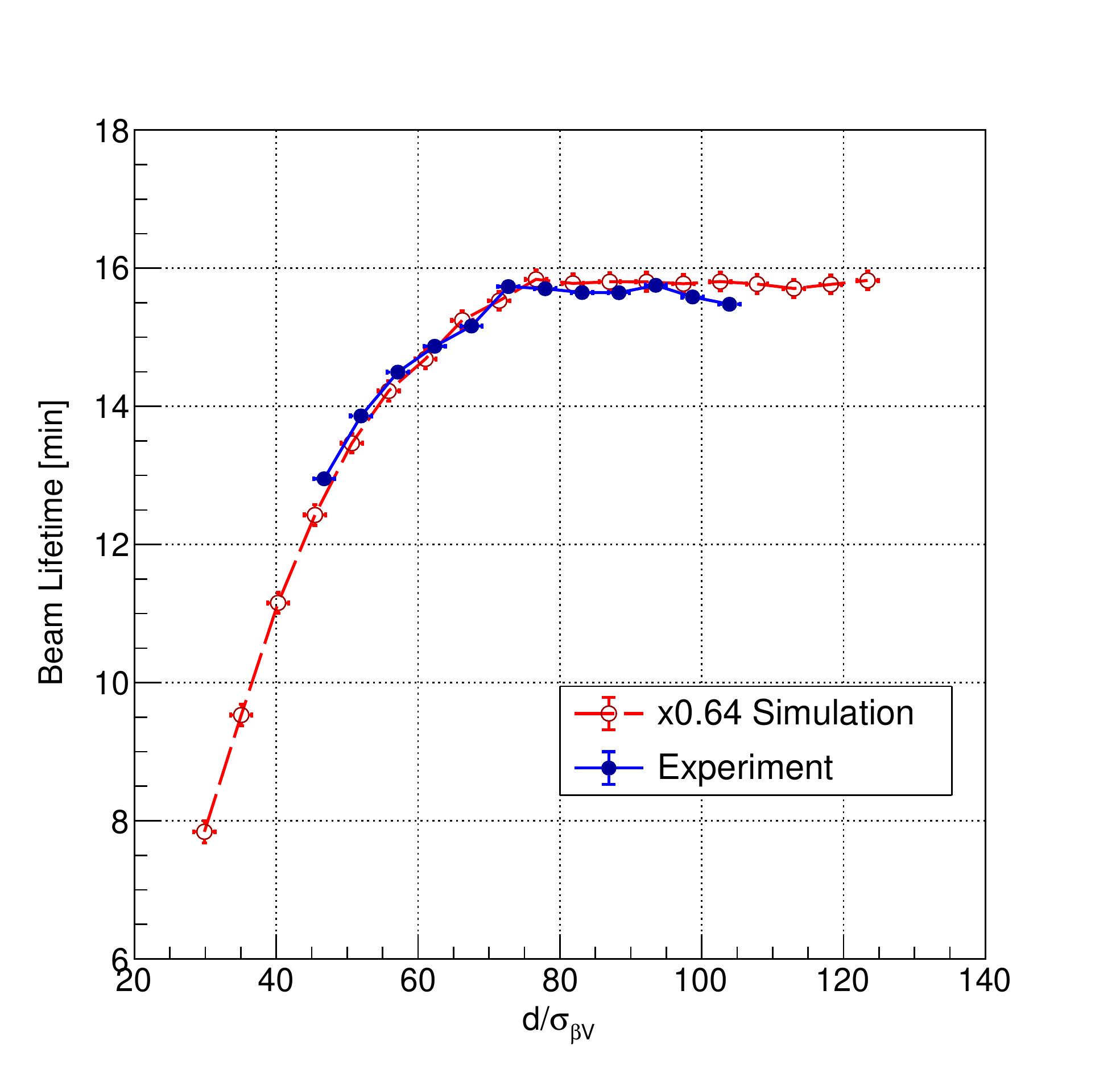}
\caption{\label{fig:fig21}LER beam lifetime versus D06V1 collimator aperture for a vertical collimator misalignment of $\rm\Delta d = \SI{0.15}{mm}$.}

\end{figure}

The measured and simulated LER beam lifetime, which reflects the integral of the beam particle loss rate around the entire ring, is shown in Figure~\ref{fig:fig21}. The simulation results were scaled to the experimental data by an empirical scale factor of 0.64 for better visualisation. The simulated trend versus D06V1 collimator aperture reproduces the experimental trend quite well, which suggests that the machine and the interaction of the beam with this collimator are modeled correctly. The need for the scale factor is not unexpected, as the SAD model of the ring does not include all detailed machine imperfections, such as misalignment of machine lattice components and noise of the current in the magnets. Our assumption that the effective atomic number of the residual gas in the beam pipe is uniform and equal to $Z_{\rm eff.} = 7$ also potentially affects the absolute scale of the simulated beam losses.

\subsubsection{\label{subsubsubsec:Tipscattering}Tip scattering}

Since a high-energy electron or positron can penetrate a high-\textit{Z} collimator head (Fig.~\ref{fig:fig11a}) and scatter without getting absorbed, the tracking code has to be able to follow tip scattered beam particles.  Figure~\ref{fig:fig11b} predicts that tungsten will impart a significant momentum and angular change to tip scattered beam particles, which makes it difficult for such particles to remain in the ring for a long time/distance. Therefore, the collimator closest to the IR is expected to be the primary source of IR backgrounds specifically due to tip scattering. 
We therefore performed an aperture scan of the LER collimator D02H4 to look for potential contributions from tip scattered particles to IR losses. As mentioned above, the horizontal collimator has an alignment uncertainty of up to \SI{1}{mm}, hence we first repeat the simulation goodness analysis, i.e. we utilize $\sum\chi^{2}/N=f(d)$ to determine the most probable horizontal misalignment between D02H4 and the center of the beam core. We obtain a horizontal offset of $\rm\Delta d = \SI{-0.4}{mm}$ for the estimated horizontal misalignment between the D02H4 collimator chamber and the nearest Q-magnet in the LER. This means that the outer collimator jaw ($d_{2}$) is closer to the beam than the inner jaw ($d_{1}$) by $\rm 2|\Delta d| = \SI{0.8}{mm}$. 

\begin{figure}[htbp]
  \subfloat{%
  \begin{tabular}{>{\centering}p{\linewidth}}
    \includegraphics[width=\linewidth]{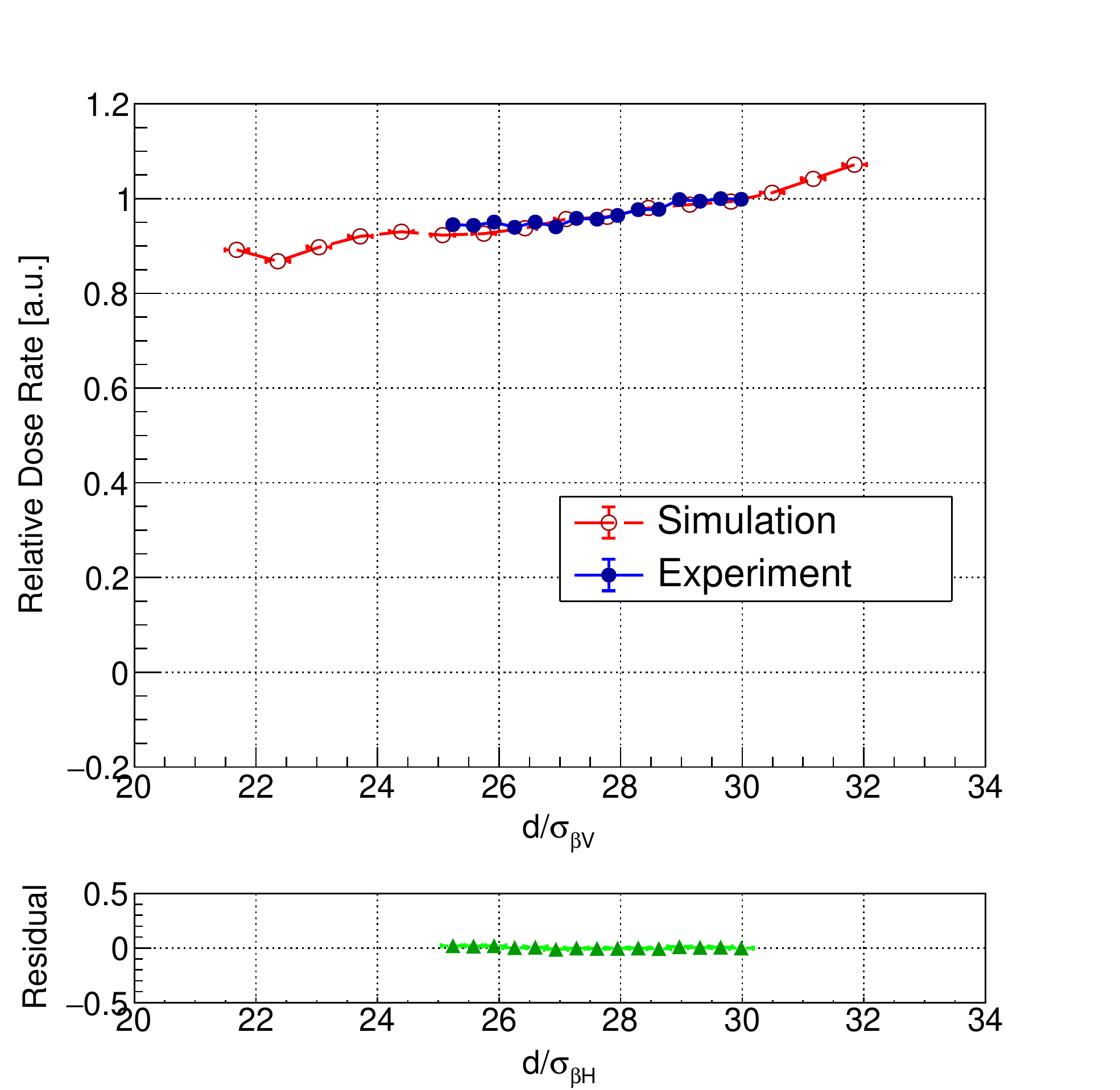}\\
    (a)~QCS-FW diamond detectors (QCS-FW-135 \& QCS-FW-225)\label{fig:fig20a}\\
  \end{tabular}
  }
  \\
  \subfloat{%
  \begin{tabular}{>{\centering}p{\linewidth}}
    \includegraphics[width=\linewidth]{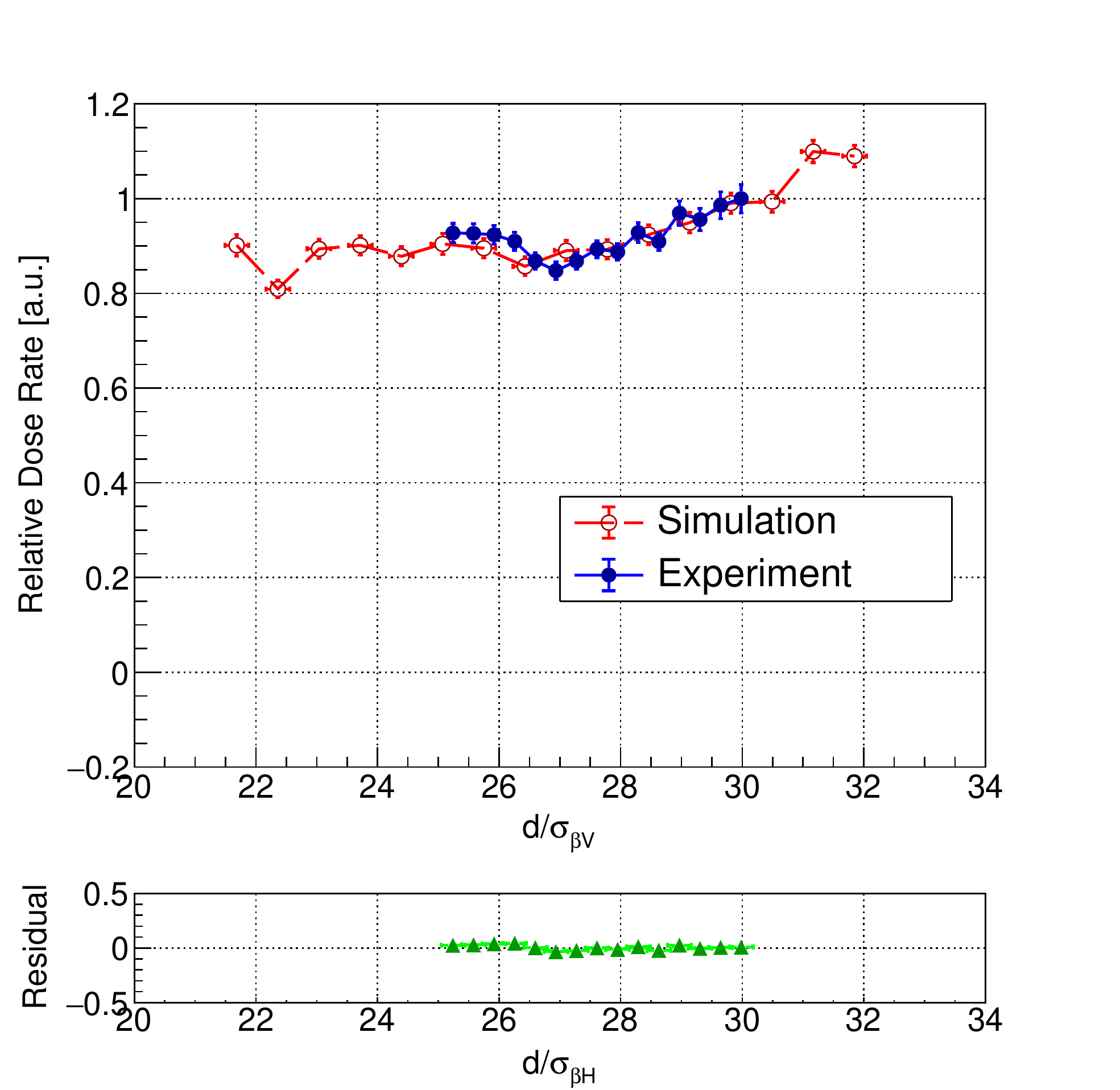}\\
    (b)~BP-FW/BW diamond detectors (BP-BW-215, BP-BW-325 \& BP-FW-215)\label{fig:fig20b}\\
  \end{tabular}
  }
  \caption{\label{fig:fig20}Simulated and observed total dose rate for two sets of diamond detectors in the interaction region versus D02H4 collimator aperture, for a simulated horizontal offset $\rm\Delta d = \SI{-0.4}{mm}$.}
\end{figure}

Figure~\ref{fig:fig20} compares the simulated and measured dose rates versus collimator aperture, where the simulation now includes the alignment offset obtained before. The measured and simulated results were normalized to have equal relative does rates in the \textit{plateau} ($29\sigma_{\rm \beta H} < d< 30\sigma_{\rm \beta H}$) and \textit{valley} ($26\sigma_{\rm \beta H} < d< 27\sigma_{\rm \beta H}$) regions. These results show that as the collimator is gradually closed, the forward QCS diamond detectors measure a decrease in the IR beam background. The beam pipe detectors, also initially observe a decrease in backgrounds. After an aperture smaller than $27\sigma_{\beta H}$ is reached, however, the beam pipe detector dose rate increases again, both in experiment and in simulation. The relative dose rate at about $25-26\sigma_{\beta H}$ is noticeable higher than the expected gradually decreasing trend of IR losses. We explain this behaviour of the background rate versus collimator aperture as due to a significant background contribution from tip scattered particles. At the narrowest measured apertures, a broader fraction of the beam halo interacts with the collimator jaws. As a consequence, a higher flux of non-absorbed particles escapes the collimator, increasing the probability that they reach Belle~II. According to the SAD simulation, most resulting beam losses are located immediately downstream of the QCS-FW detectors. Therefore, these diamond detectors are not very sensitive to this relatively small effect, while the BP-FW/BW diamond detectors do detect this phenomenon.

\begin{figure}[htbp]
\centering
\includegraphics[width=\linewidth]{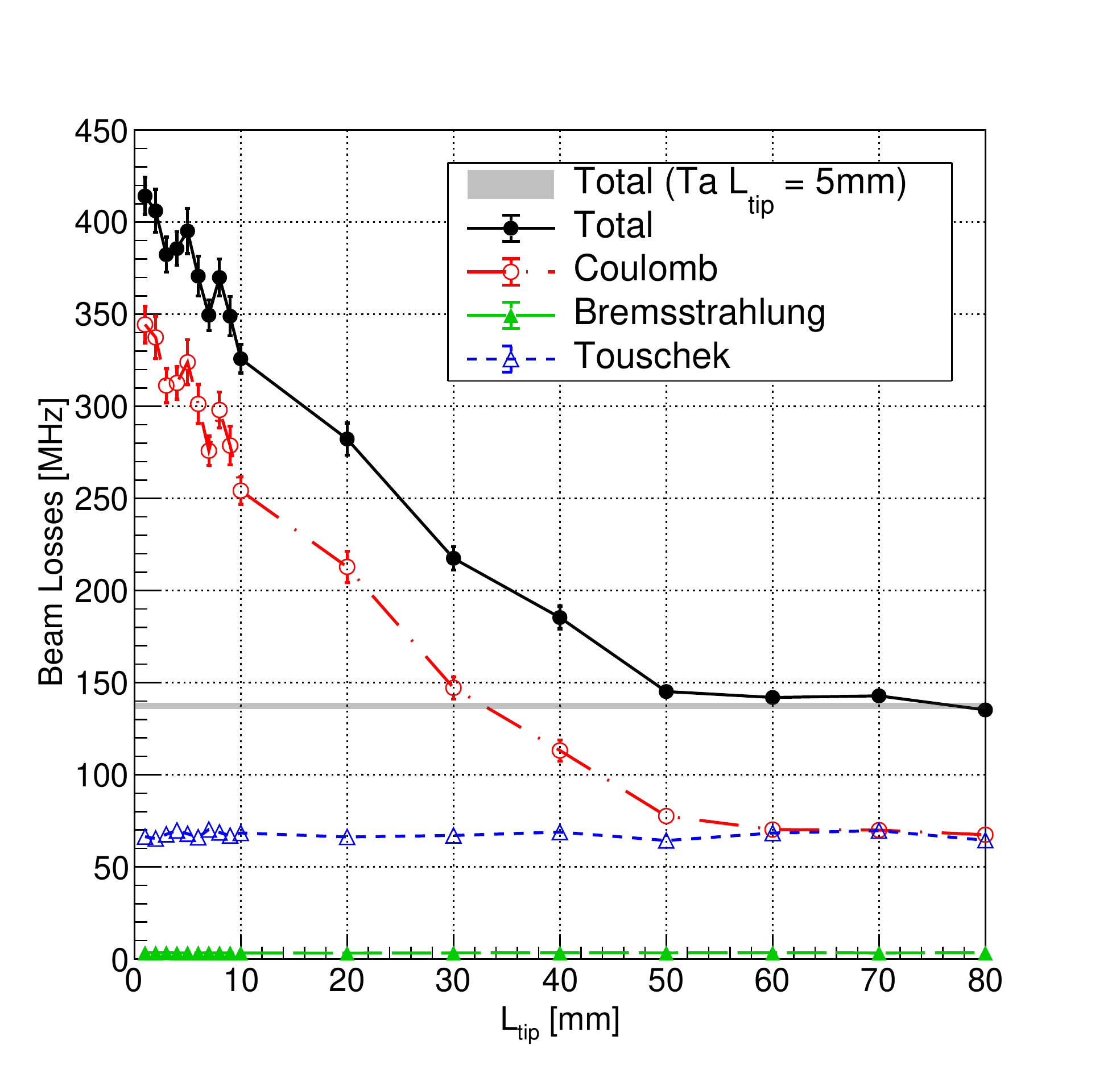}
\caption{\label{fig:fig22}Example results from a collimator optimization that utilizes the newly implemented tip scattering simulation. The solid, black line shows expected total single-beam IR loss rate versus D06V1 collimator head length, for a head made of graphite. For comparison, the grey horizontal line shows results for a \SI{5}{mm} tantalum head.}

\end{figure}

Our addition of tip scattering to the background simulation has already proven useful when deciding on future collimator designs and configurations. One example is the September 2020 replacement of the D06V1 tantalum collimator head with graphite. The initial design length of the graphite collimator tip was \SI{60}{mm}. We performed a set of simulations with different lengths ($L_{\rm tip}$) to see if this dimension could be reduced, to minimize the impedance of the collimator~\cite{REF25}. Figure~\ref{fig:fig22} shows predicted beam loss rates versus $L_{\rm tip}$. It can be seen that for $L_{\rm tip} > \SI{50}{mm}$ the effect of the graphite head is almost the same as for a tantalum head of \SI{5}{mm} length (grey horizontal line). This outcome justifies the choice of \SI{60}{mm} for graphite. Unfortunately, for $L_{\rm tip} < \SI{50}{mm}$ IR losses start to increase due to tip scattered particles. The induced momentum and angle changes are insufficient to kick scattered particles out of the machine's aperture before they reach the interaction region. A shorter head leads to higher Coulomb scattering losses, as the vertical collimator does not significantly affect the Touschek halo, which is distributed mainly in the horizontal plane. As a result, the D06V1 graphite head length cannot be shorter than about \SI{50}{mm} if we wish to keep the same level of LER backgrounds as with a \SI{5}{mm} tantalum head.

\section{\label{sec:SummaryAndConclusions}Summary and conclusions}

As SuperKEKB improves further on its world record luminosity, continued simulation and mitigation of beam-induced backgrounds will be essential. Collimators play a crucial role in Belle~II background mitigation and in cleaning up the halo of the circulating beams. To find an optimal collimator configuration and to predict beam particle losses in the machine, an extended and improved Monte-Carlo simulation framework was developed. This version includes more realistic shapes of all collimator jaws, a new sequential tracking procedure that enables off-line collimator scans and optimization, and a model for the beam particle scattering in the collimator material. Several other crucial improvements were recently implemented, for example position-dependent simulation of measured vacuum pressure in each ring, but these will be reported separately. The new simulation framework enables optimization of collimator settings in a few hours, while the old framework required several days of CPU time. 

The described implementations and improvements of the accelerator particle tracking result in greatly improved agreement with experimental data. Background simulation validation with Belle~II sub-detectors will be reported separately in the future. Here, we have focused on validating the effect of collimators in the simulation, as this is where we found the largest discrepancies with experiment. Simulation of LER collimator aperture scans reproduce measurements well, confirming the newly implemented collimator shapes and tip scattering models. These scans are sensitive to misalignments of collimator chambers with respect to the beam center. The presented method of utilizing the matching to simulation may thus be used for improved collimator alignment in the future. The findings also emphasize the importance of more precise positioning of accelerator components.

Finally, the enhanced simulation framework also allows us to predict the effect of new machine elements and to study possible improvements for further background reduction in the future.

\begin{acknowledgments}
We enthusiastically thank the SuperKEKB commissioning group for operation of the accelerator; the KEK computing group for on-site support; the Belle~II diamond detector group for assistance with dose rate calibrations; J.~Schueler and T.~E.~Browder~(University of Hawaii) and L.~Lanceri~(INFN, Sezione di Trieste) for their feedback on the manuscript.  This work was supported by the U.S. Department of Energy (DOE) via Award Number DE-SC0007852 and via U.S. Belle~II Operations administered by Brookhaven
National Laboratory (DE-SC0012704).
\end{acknowledgments}.
\bibliography{bibliography}

\end{document}